\documentclass[11pt, a4paper, oneside]{article}
\usepackage[body={16.5cm, 21.0cm},left=1.5cm,right=1.5cm]{geometry}
\usepackage{natbib}
\usepackage{amsmath, amssymb, amsthm}
\usepackage{amsfonts}
\usepackage{graphicx}
\usepackage{dsfont} 
\usepackage{color}
\usepackage{nicefrac}
\usepackage{multirow} 
\RequirePackage[colorlinks,citecolor=blue,urlcolor=blue]{hyperref}

\usepackage{accents}
\usepackage{setspace}
\usepackage{pifont}
\usepackage{utopia}

\usepackage{fancyhdr}
\usepackage{soul}
\numberwithin{equation}{section}

\usepackage{tikz,xcolor}
\definecolor{lime}{HTML}{A6CE39}
\DeclareRobustCommand{\orcidicon}{%
	\begin{tikzpicture}
	\draw[lime, fill=lime] (0,0) 
	circle [radius=0.16] 
	node[white] {{\fontfamily{qag}\selectfont \tiny ID}};
	\draw[white, fill=white] (-0.0625,0.095) 
	circle [radius=0.007];
	\end{tikzpicture}
	\hspace{-2mm}
}

\foreach \x in {A, ..., Z}{%
	\expandafter\xdef\csname orcid\x\endcsname{\noexpand\href{https://orcid.org/\csname orcidauthor\x\endcsname}{\noexpand\orcidicon}}}

\theoremstyle{plain}

\newtheorem{theorem}{Theorem}[section]
\newtheorem{proposition}{Proposition}
\newtheorem{lemma}[theorem]{Lemma}

\theoremstyle{remark}

\newcommand{\keywords}[1]{{\scriptsize \noindent \textbf{KEY WORDS AND PHRASES:}} {#1}}
\newcommand{\msc}[1]{{\scriptsize \noindent \textbf{MSC2020 SUBJECT CLASSIFICATIONS:}} {#1}}

\newenvironment{pfofThm}{\noindent{\bf Proof of Theorem}}{\hfill $\square$ \\}
\newenvironment{pfofLem}{\noindent{\bf Proof of Lemma}}{\hfill $\square$ \\}

\RequirePackage[utf8]{inputenc} 
\RequirePackage[T1]{fontenc}    
\RequirePackage{multicol,booktabs,longtable}       
\RequirePackage{bm}  
\RequirePackage{nicefrac}       
\RequirePackage{microtype}      
\RequirePackage{cleveref}       
\RequirePackage{psfrag,epsf}
\RequirePackage{caption,subcaption}
\usepackage{enumitem}

\def\E{\mathbb{E}}
\def\R{\mathbb{R}}

\def\Z{\mathbb{Z}}
\def\N{\mathbb{N}}

\newcommand{\Hcal}{\mathcal{H}}

\newcommand{\Lcal}{\mathcal{L}}
\newcommand{\Mcal}{\mathcal{M}}
\newcommand{\Gcal}{\mathcal{G}}
\newcommand{\Fcal}{\mathcal{F}}
\newcommand{\Kcal}{\mathcal{K}}
\newcommand{\Scal}{\mathcal{S}}
\newcommand{\Tcal}{\mathcal{T}}

\newcommand{\abs}[1]{\left\lvert #1 \right\rvert}
\newcommand{\norm}[1]{\left\Vert #1 \right\Vert}
\newcommand{\dash}{^{\prime}}

\DeclareMathOperator*{\argmin}{arg\,min}

\newcommand{\bb}[1]{\boldsymbol{#1}}
\newcommand{\bbhat}[1]{\widehat{\boldsymbol{#1}}}
\newcommand{\tr}{^{\intercal}}

\newcommand{\kernel}[1]{\bb{K}(\bb{x}, \bb{X}(#1))}

\newcommand{\prob}{\mathbb{P}}
\newcommand{\normdist}{\mathcal{N}}
\newcommand{\cov}{\mathrm{Cov}}
\newcommand{\var}{\mathrm{Var}}

\newcommand{\conv}[1]{\ensuremath{\, \displaystyle {\mathop {\longrightarrow}^{#1}}}\, }

\title{Nonparametric quantile regression for spatio-temporal processes}

\author{
 {Soudeep Deb}\orcidA{}\\  {\small Indian Institute of Management Bangalore}\\
  \and {Cl\'{a}udia Neves}\orcidB{}\\ {\small King's College London}\\
  \and {Subhrajyoty Roy}\orcidC{}\\  {\small Indian Statistical Institute Kolkata}\\
}%
\date{}

\begin{document}

\maketitle


\abstract{In this paper, we develop a new and effective approach to nonparametric quantile regression that accommodates ultrahigh-dimensional data arising from spatio-temporal processes. This approach proves advantageous in staving off computational challenges that constitute known hindrances to existing nonparametric quantile regression methods when the number of predictors is much larger than the available sample size. We investigate conditions under which estimation is feasible and of good overall quality and obtain sharp approximations that we employ to devising statistical inference methodology. These include simultaneous confidence intervals and tests of hypotheses, whose asymptotics is borne by a non-trivial functional central limit theorem tailored to martingale differences. Additionally, we provide finite-sample results through various simulations which, accompanied by an illustrative application to real-worldesque data (on electricity demand), offer guarantees on the performance of the proposed methodology.}

\vspace{0.2cm}

\keywords{Asymptotic theory, electricity demand forecasting, gaussian processes, multivariate quantiles.}
\vspace{0.2cm}

\msc{Primary 62M20, 62A99; secondary  60G15, 60G70.}



\section{Introduction}\label{sec:introduction}

Within the context of various economic, environmental and societal research problems, lies the common aim of wanting to estimate the effect of a covariate on the behavior of the quantiles of a continuous response variable whilst it is also well known that a simple regression model explaining the conditional expectation of the response variable may turn out short in providing a satisfactory description of the full spread of the targeted conditional distribution, particularly at the level of tail-related data  \citep[cf.][]{reich2012spatiotemporal}. 

Originally proposed by \cite{koenker1978regression}, even in its most simplified setting, quantile regression serves as a valuable alternative  to the familiar regression on conditional expectation, for it is not as heavily reliant on the typical assumption about normally distributed errors for assessing significance of the postulated model. Indeed, working off the inherent nonparametric nature of conditional quantiles, quantile regression proves as a reliable inference methodology to the extent of readily accommodating heavy-tailed distributions for the errors, also being robust to the presence of outliers \citep[see][]{chaudhuri1997average,ezzahrioui2008asymptotic}. In more complex situations, especially those concerning time series modeling or involving space-time changing attributes, quantile regression recognizably offers a plethora of opportunities to statistically meaningful analyses of empirical data from real-life applications.

The leading example in this paper has dedicated focus on conditional quantiles of electricity load profiles obtained from smart-meter records installed in many UK households. The key motivation behind this illustrative application is to gain improved understanding of peak-load events so as to help distribution network providers maintain a stable supply chain at particular times of the day or week and ward off rare but sometimes inevitable power outages \citep[][]{jacob2020forecasting}. Like with many other fields in applied science, an important difficulty arises from the dimension of the space-time covariate (thus infinite-dimensional in theory) becoming increasingly large as the sample size increases. We tackle this problem through a multivariate time series approach under minimally restrictive assumptions on the data generating process, hence making it amenable to translate to other applied settings given the  added flexibility for adjusting to different factors and confounding effects. Before delving into both rationale and technical aspects of our proposed suite of statistical methods concerning quantile regression on a spatio-temporal dimension, a succinct account of extant literature in this context is pertinent at this point. 

There has been primarily two distinct nonparametric approaches towards modeling conditional quantiles when dealing with a response variable at a time. One such approach hinges on a pinball type loss function conditional on the value of the covariate and views the resulting estimation problem from the optimization modeling paradigm. In order to translate this into a viable and tractable problem, however, one often needs to make a set of assumptions which have typically ranged from a continuous functional of the covariate of interest, to functionals belonging to a reproducible kernel Hilbert space \citep{takeuchi2006nonparametric}, and to polynomial functions of the covariate \citep{zhou2009local}, all of which resulting in smooth quantile curves. A common caveat to this type of approaches is that they can still suffer from the well-known quantile crossing effect, which will lead to extra constraints yet to be factored into the optimization setup. To combat this issue, a nonparametric kernel estimation of the conditional distribution function was put forward \citep[cf.][and references therein]{ferraty2005conditional,ezzahrioui2008asymptotic2} which, upon a relatively straightforward inversion procedure, has had proven efficacy in  yielding conditional quantile curves that satisfy the non-crossing criterion. However, because there is a necessary inversion step involved in the latter, the former challenge now shifts to the smoothness property of the estimated quantile curve which is not always attainable in practice, thus posing a critical hurdle to the modeling of real-worldesque data.

When looking at multivariate time series or spatio-temporal data, a case which this paper seeks to address, we note that despite the vastly accumulating literature on conditional quantile function estimation in the independent setting, the development of flexible statistical methodology for the underpinning stochastic processes is still relatively in its infancy. Progress in these domains has been limited, possibly by the challenges faced when accounting for the core dependency structure in a way that does not break down at the laws of physics governing the process generating the data and therefore is capable of producing realistic inference. To the best of our knowledge, only a few successful attempts in this direction have been made thus far. A couple of them \citep[e.g.][]{reich2011bayesian,das2017analyzing} are Bayesian approaches that tend to anchor on basis polynomials, with ensuing considerations in eliciting prior distributions to the relevant parameters alongside discussion on which allowances must be brought into practice to make the method feasible. Because any misspecification involving the latter is detrimental to the overall quality and feasibility of inference, the possibility of undermining end-users' confidence in the statistical methods being employed cannot be wholly discarded. The approach taken by \cite{reich2012spatiotemporal,duan2021spatio}, among others, considers each quantile curve at different locations separately, with the levels of the quantile curves not affecting each other except for the non-crossing condition. However, this implies that such approaches may fail in the situations where observations at most locations follow a general trend (thus remaining close to the median) with a few remainder locations exhibiting disparate behavior from the bulk. 

In this paper, we introduce a nonparametric estimator of the conditional quantile of a continuous response variable. Our statistical procedure has the advantage of being able to deliver smooth estimates whilst demonstrating to stave off the above-mentioned quantile crossing effect. The extremity issue is overcome by modeling the multivariate quantiles of the relevant distribution through an appropriate random element of the unit euclidean ball. The asymptotic properties are developed as part of the theoretical study of the proposed statistical methodology, both aimed at estimation and testing. We consider the asymptotic framework where both the number of time-points and the number of locations grow to infinity, which required significant adaptation of the conventional delta-method used for deriving consistency and asymptotic normality of estimators, since our approach leverages the sub-Gaussian properties of the data to maintain higher-order moments under control for ensuring tightness. Moreover, ensuring consistency has dictated that weak convergence of the process must hold in a uniform sense in order to account for the changing vector spaces with varying dimension, hence being the source of additional complexity. This was handled by employing concentration inequalities based on a sequence of compact sets for guaranteeing boundedness in the differences between estimated and true quantiles. Lastly, to prove asymptotic normality, we make use of Portmanteau's theorem and show that, for a rich class of smooth functions with bounded derivatives, the expectation of the desired functional of the normalized estimators converges to the expectation of the same functional value applied on the appropriate slice of Gaussian process, which we also identify.

The structure of this paper is as follows. The statistical methodological details and the key results developed are described in Sections \ref{Sec:MainAssump}, \ref{Sec:Estimator} and \ref{Sec:Asymptotics}. Following the exposition of asymptotic theory, we go on to developing simultaneous confidence bands for the spatio-temporal quantile curves across all locations and time-points in Section \ref{sec:ht}. Associated statistical tests of hypotheses are also described therein. In order to maintain the flow of the paper, we defer the proofs of the theoretical results until Section~\ref{sec:main-proofs}. In Section \ref{Sec:real-data}, we present our maiden application of the proposed estimator to electricity load forecasting based on smart-meter data collected as part of the Thames Valley Vision project \citep[details in][]{jacob2020forecasting}. This section focuses on inference, encompassing estimation, prediction and hypothesis testing, with a particular emphasis on two testing procedures for homogeneity. Numerical experiments that provide empirical evidence supporting the validity of our statistical methods are conducted as a supplement to this study. Notably, the fundamental large sample properties elucidated in the paper are mirrored in the finite sample simulation study detailed in the supplementary material accompanying this paper. 

\section{Spatio-temporal framework for quantile regression}
\label{Sec:MainAssump}

This section presents the framework for the proposed quantile regression for spatio-temporal processes, also furthering related works by \cite{de2006multivariate,zhou2010nonparametric,deb2021nonparametric}.

Let $\left\{Y(t, s):\, t \in \Tcal, s \in \Scal \right\}$ be a spatio-temporal process indexed by temporal domain consisting of a (dense) temporal horizon $\Tcal = [0, 1]$ and by a countable spatial horizon set $\Scal \subset \R^2$. For all $d\in \mathbb{N}$, let $\bb{X}: \Tcal \rightarrow \chi \subseteq \R^d$ be a $d$-dimensional temporal covariate process such that for each $t \in \Tcal$, $\bb{X}(t) = \left(X_1(t), \ldots, X_d(t) \right)'$ is a predictor variable in $\chi \subseteq \R^d$. In the sliding scale of generality, this covariate process can be allowed to be made fairly generic so as to encapsulate external factors that might affect the temporal dependence to the main process $Y(t, s)$, to being set out as specialized into a given periodic component, for example, aiming to capture inherent seasonality. Henceforth , our interest lies in the conditional space-time process  $Y(t, s) \mid \{ \bb{X}(t) : t \in \Tcal\}$ with expectation and covariance structure respectively denoted by
\begin{equation}
      \E(Y(t, s)) = \mu(\bb{X}(t), s) \quad  \mbox{ and }  \quad
      \cov(Y(t, s_1), Y(t, s_2)) = \Sigma(\bb{X}(t), s_1, s_2),
    \label{eqn:general-setup}
\end{equation}
for every $s, s_1, s_2 \in \Scal$ and $t \in \Tcal$. Note that the temporal dependence in the process $Y(t, s)$ is essentially driven by that of the associated covariate process $\bb{X}(t)$. Of course, it would be beguiling to assume that $Y(t, s)$ can actually be observed in the way of covering the whole set $\Tcal \times \Scal$, so we must settle with $Y(t, s)$ indexed by the discrete subset of time points $\Tcal_n = \{t_1, t_2, \dots t_n \}$ and locations $\Scal_p = \{ s_1, s_2, \dots s_p \}$, where both $n$ and $p$ are assumed sufficiently large, i.e.\ both tending to $\infty$ albeit in a controlled manner for the latter which we impose from the outset as an intermediate sequence $p=p_n$ such that $p/n \rightarrow 0$. This renders a more realistic framework for inference with a practical resonance: for each $t \in \Tcal_n$, suppose that the data is generated according to the nonparametric spatial regression model:
\begin{equation}\label{eqn:setup}
    \bb{Y}(t) = 
    \begin{pmatrix}
    Y(t,s_1)\\ Y(t, s_2) \\ \vdots \\ Y(t, s_p)
    \end{pmatrix} 
    = \bb{\mu}(\bb{X}(t)) 
    + \bb{\Sigma}^{1/2}(\bb{X}(t)) \begin{pmatrix}
   \xi_{t1}\\ \xi_{t2} \\ \vdots \\ \xi_{tp}
    \end{pmatrix},
\end{equation}
where $\bb{\mu}(\bb{X}(t))$ is a random vector in $\mathbb{R}^p$ of the form
\begin{equation*}
    \bb{\mu}(\bb{X}(t)) := \begin{pmatrix}
      \mu(\bb{X}(t), s_1)\\
      \mu(\bb{X}(t), s_2)\\
      \vdots \\
      \mu(\bb{X}(t), s_p)
    \end{pmatrix},
\end{equation*}
the covariance matrix is given by $\bb{\Sigma}(\bb{X}(t))$, whose $(i, j)$-th entry is $\Sigma(\bb{X}(t), s_i, s_j)$ for $1 \leqslant i, j\leqslant p$, and the (unobserved) errors $\left(\xi_{tj}\right)_{j=1, \dots p}$ are independent and identically distributed with mean zero, positive density on an interval containing $0$ (albeit not necessarily that of a normal distribution) and unit variance. Hereafter, $\bb{\xi}(t)$ denotes the vector of $\xi_{tj}$'s. 

The framework~\eqref{eqn:setup} lays bare that the key problem of estimating a spatial quantile at any given time $t \in \mathcal{T}$ fits well into the logic of estimating a $p$-variate quantile for the distribution of $\bb{Y}(t)$ centered at time-points $t_1, t_2, \hdots, t_n$. It remains quite general on the whole, and especially in two ways outlined as follows. Firstly, it accommodates multivariate time series data even when there is no inherent spatial structure in the coordinates but rather a rank correlation can be inferred from domain knowledge, such as daily closing prices of multiple stock indices. If relevant information about the spatial dependency structure is available \emph{a priori}, it can be encapsulated in $\bb{\Sigma}(\cdot)$ through a set of suitable constraints. These constraints range from being precisely defined, fundamentally reliant on a certain postulated parametric model, to being defined by a given class of (unknown) distributions in a manner that echoes a familiar balancing act: the robustness-efficiency trade-off. Secondly, such a formulation enables inference using multivariate quantiles that may describe different degrees of extremity on different coordinates as it allows for possible different tail behavior at nearby locations. This justifies the focus on nonparametric estimation of the conditional multivariate quantiles of $\bb{Y}(t)$, given the temporal covariate process $\bb{X}(t)$, assumed predictable.



To this date, various definitions for a quantile of a multivariate distribution have been put forward in the literature, notably those drawing on convex hulls \citep{eddy1985ordering}, minimum volume sets \citep{einmahl1992generalized}, geometric generalization of spatial median \citep{chaudhuri1996geometric} and center outward quantile functions \citep{figalli2018continuity}. Among these, the geometric concept of a quantile in $\mathbb{R}^d$, $d \geqslant 2$, originating from \cite{chaudhuri1996geometric} and endorsed in \cite{KonenPaindaveine22}, stands out as the most closely related to the present paper.

Indeed, central to this paper is the loss function $\Phi(\bb{u}, \bb{v}) = \norm{\bb{v}} + \langle \bb{u}, \bb{v}\rangle$ ($\norm{\bb{v}}$ denotes the usual Euclidean norm of $\bb{v}$), for all $\bb{u} \in B^{d-1} = \{ \bb{x} \in \R^d : \norm{\bb{x}} < 1 \}$ and the estimated quantile $\bb{v} \in \R^d$, in which setting a recognizable form of the  multivariate $\bb{u}$-quantile of a $d$-variate random vector $\bb{X}$ beckons: 
\begin{equation}\label{eqn:q-simple}
    Q(\bb{u}) := \arg\min_{\bb{q} \in \R^d} \, \E\left[ \Phi(\bb{u}, \bb{X} - \bb{q}) \right],
\end{equation}
\citep[cf.][]{ChowChau19}. As desired, a choice of $\bb{u} \in B^{d-1}$ in~\eqref{eqn:q-simple} towards the boundary, i.e. $\norm{\bb{u}} \to 1$, leads to a (high or low) extreme quantile, whereas fixing $\bb{u}$ near the origin yields spatially centered measurable functionals, including expectiles such as the mean vector ascribed to the $\Lcal_2$-objective function in \eqref{eqn:q-simple} or the spatial median in connection with $\Lcal_1$-norm. Since the vector $\bb{u}$ depends on the dimension $p$ of the covariate, and  $p=p_n$ is such that $p\rightarrow \infty$ as $n \rightarrow \infty$ (in a controlled manner stipulated before), we shall consider a spatial process $u$ (measurable with respect to the Lebesgue measure on $\R^p$) satisfying
\begin{equation*}
    \int_{\Scal} \vert u(s) \vert^2 ds \leqslant 1,
\end{equation*}
which, for every finite restriction $\Scal_p$ of $\Scal$, gives rise to $\bb{u}\vert_{\Scal_p} = (u(s_1), \dots u(s_p)) \in B^{p-1}$. In a similar vein, we now define for every $t \in \Tcal, \bb{x} \in \chi$, the $\bb{u}$-conditional spatial quantile of $\bb{Y}(t)$ given $\bb{X}(t) = \bb{x}$ as
\begin{equation}\label{eqn:parameter}
    Q(\bb{u} ,\bb{x} )\vert_{\R^p} = \arg\min_{\bb{q}\in \R^p} \, \E\left[ \Vert \bb{Y}(t) - \bb{q}\Vert +  \langle \bb{u}\vert_{\Scal_p}, \bb{Y}(t) - \bb{q} \rangle \mid \bb{X}(t) = \bb{x}\right],
\end{equation}
where $Q(\bb{u} ,\bb{x})\vert_{\R^p}$ highlights that the conditional quantile process $Q(\bb{u} ,\bb{x})$ is indexed by a $p$-dimenstional spatial domain $\Scal_p$. The set $\Scal_p$ has to be asymptotically dense in $\Scal$ as $p \rightarrow \infty$, i.e., for any $s \in \Scal$ and however small $\epsilon > 0$, we can find a point $s^\ast \in \Scal_p$ within the ball of radius $\epsilon$ centered at $s$ for sufficiently large $p$.  It is important to stress that the inclusion of the temporal domain adds to a body of contemporary works in a substantive way since, to the extent of our knowledge, research thus far refrained to the study of nonparametric spatial quantile regression. Our proposal is distinctive in that the criterion (loss) function presiding the M-estimation contemplated in \eqref{eqn:parameter} is no longer a fixed function, but rather it is allowed to vary with $p$ whilst preserving many of the fundamental properties for quantile regression investigated in \cite{ChowChau19}. This is the key insight to expanding the notion of a spatial (multivariate) quantile across a direction, furthering the aim of tackling the spatio-temporal regression estimation problem stated in~\eqref{eqn:parameter}. The precise meaning of this will become more apparent in Section~\ref{Sec:Estimator}.

In this instance, a natural question that comes to the fore is whether such a definition of conditional quantile process (as in~\eqref{eqn:parameter}) entails identifiability across the set $\Scal_p$. For example, if $\Scal^\ast \subseteq \Scal$ is a superset of $\Scal_p$ for some fixed $p$, then the restriction of $Q(\bb{u}, \bb{x})\vert_{\Scal^\ast}$ to the set of points $\Scal_p$ should be same as $Q(\bb{u}, \bb{x})\vert_{\R_p}$. To see this, we consider $\Scal^\ast = \Scal_p \cup (\Scal^\ast - \Scal_p) = S_1 \cup S_2$ and correspondingly define $\bb{Y}_1(t), \bb{Y}_2(t)$ and $\bb{q}_1, \bb{q}_2$ as the restrictions of $\bb{Y}(t)$ and $\bb{q}$ respectively. Finally, we note that
\begin{multline*}
    Q(\bb{u}, \bb{x})\vert_{\Scal^\ast} = \argmin_{\bb{q}_1, \bb{q}_2} \E\left[ \sqrt{\Vert \bb{Y}_1(t) - \bb{q}_1\Vert^2 + \Vert \bb{Y}_2(t) - \bb{q}_2\Vert^2 } \right. \\
    \left. + \langle \bb{u}\vert_{S_1}, \bb{Y}_1(t) - \bb{q}_1 \rangle + \langle \bb{u}\vert_{S_2}, \bb{Y}_2(t) - \bb{q}_2 \rangle \mid \bb{X}(t) = \bb{x} \right].
\end{multline*}
\noindent It now clearly follows that optimizing with respect to $\bb{q}_1 \in \R^p$ yields a minimizer of \eqref{eqn:parameter} for any fixed value of $\bb{q}_2$, thus ensuring identifiability.

%
\section{Nonparametric spatio-temporal quantile estimation}
\label{Sec:Estimator}

For ease of introduction, we first address the case where we look at one location at a time. This corresponds to setting $p = 1$ in the basic model~\eqref{eqn:setup}, whereby the canonical nonparametric quantile estimator gives rise to the kernel-weighted version of the quantile $q$ of order $\tau \in (0,1)$ of the corresponding empirical probability distribution function of $Y(t)$ given $\bb{X}(t) = \bb{x}$ on $\chi= \{x_1, \ldots, x_n\} \subset \R^d$. Formally, for $u\equiv u(\tau) = (2\tau - 1)$, $0 < \tau < 1$, the quantile $q$ of order $\tau$ is the minimizer of 
\begin{equation}\label{EstimatorSimple}
    \mathcal{M}_{u,n}^{(1)}(q)= \sum_{i=1}^n k(\bb{x}, \bb{X}(t_i)) \left( \vert Y(t_i) - q\vert + u \cdot (Y(t_i) - q) \right),
\end{equation}
\noindent where $k(\cdot, \cdot)$ is a suitably chosen kernel function. In this respect, we note that $\mathcal{M}_{u,n}^{(1)}$ is convex on $\R$. With $\tau= 1/2$ standing for the median as the minimizer of $\mathcal{M}_{u,n}^{(1)}$, $u$ can be viewed as a gauge for central displacement or tail-relatedness, since the greater the $\vert u\vert$ the more tail-relatable the quantile. Clearly, finding minimizers of~\eqref{EstimatorSimple} configures an $M$-estimation problem, and with the stated sample criterion function incorporating a kernel function of bandwidth $b_n$, this is one of a nonparametric nature. By this token, a generalization to the $p$-multivariate case can be readily achieved by taking suitably normalized matrix-valued kernel functions $\bb{K}(\cdot, \cdot)$ into the inner product that constitutes~\eqref{EstimatorSimple}, thus giving way to the following definition of a nonparametric spatio-temporal $\bb{u}$-quantile that mirrors \eqref{eqn:parameter} in the associated sequence of statistical experiments: for $\bb{u}\vert_{\Scal_p} \in \R^p$ comprised of the set $\{ u(s): s \in \Scal_p \}$, define the estimator 
\begin{equation} \label{eqn:estimator}
    \hat{\bb{q}}_n(\bb{u} ,\bb{x}) := \arg\min_{\bb{q}\in \R^p} \mathcal{M}_{\bb{u},n}^{(p)}(\bb{q}),    
\end{equation}
where 
\begin{equation}
    \mathcal{M}_{\bb{u},n}^{(p)}(\bb{q}):= \frac{1}{n} \sum_{i=1}^n  \left\{\left\Vert \bb{Y}(t_i) - \bb{q}\right\Vert_{\bb{K}(t_i)} + \left\langle\bb{u}\vert_{\Scal_p}, \bb{Y}(t_i) - \bb{q} \right\rangle_{\bb{K}(t_i)} \right\},
    \label{eqn:m-u-n}
\end{equation}
for which we write
\begin{align*}
    \bb{K}(t) & := \kernel{t}\\
    \left\Vert \bb{Y}(t) - \bb{q}\right\Vert_{\bb{K}(t)}^2 
    & := \norm{\bb{K}(t)(\bb{Y}(t) - \bb{q})}^2,\\
    \left\langle\bb{u}, \bb{Y}(t) - \bb{q} \right\rangle_{\bb{K}(t)} 
    & := \bb{u}\tr \bb{K}(t) (\bb{Y}(t) - \bb{q}).
\end{align*}
\noindent As the objective function $\mathcal{M}_{\bb{u}, n}^{(p)}$ in \eqref{eqn:m-u-n} has been highlighted to depend on both the parameter dimension $p$ and sample size $n$, in view of~\eqref{eqn:parameter} its counterpart population version (for criterion function) is defined as
\begin{equation*}
    \mathcal{M}_{\bb{u}}^{(p)}(\bb{q}) := \E\left[ \Vert \bb{Y}(t) - \bb{q}\Vert +  \langle \bb{u}\vert_{\Scal_p}, \bb{Y}(t) - \bb{q} \rangle \mid \bb{X}(t) = \bb{x}\right].
\end{equation*}
In Section~\ref{Sec:Asymptotics}, this setting will be expanded with $p$ taken as a sequence of positive integers indexed in $n$.

%
Before going any further, we spell out the key assumptions underpinning the basic asymptotic results established in this paper; each of which will be called upon as deemed appropriate and/or necessary throughout.

\begin{enumerate}[label=(A\arabic*),ref=(A\arabic*)]
    \item\label{assum:predictable} The covariate process $\bb{X}(t)$ is a predictable process, i.e., there exists a sequence of random variables $\{ \epsilon_{t}: t = t_1, t_2, \dots \}$, with its natural filtration $\Fcal_{t}$ independent of the random variables $\xi_{tj}$, such that $\bb{X}(t_i)$ is $\Fcal_{t_{(i-1)}}$-measurable for any $t_i \in \Tcal_n$.
    
    \item\label{assum:openconvex} There exists a sequence of convex, open sets $\mathcal{U}_1 \subseteq \R, \mathcal{U}_2 \subseteq \R^2, \ldots, \mathcal{U}_p  \subseteq \R^p$ where the sequence of criterion functions $\mathcal{M}_{\bb{u},n}^{(p)}$, defined as in~\eqref{eqn:m-u-n} takes on the vector-valued $\bb{q} \in \mathcal{U}_p$, for all $t_i \in \Tcal_n$, and for any fixed $p \in \mathbb{N}$ not depending on $n$. Moreover, we assume that $\inf_{\bb{q} \in \mathcal{U}_p} \inf_{t_i \in \Tcal_n} \Vert \bb{Y}(t_i) - \bb{q}\Vert > \delta$ for sufficiently small $\delta > 0$, independent of $n$ and $p$. 
    
    \item\label{assum:kernelmatrix} The kernel matrix $\kernel{t}$ is a positive definite ($p\times p$) matrix for all $t \in \Tcal$ and $\bb{x} \in \chi$, and the support of the conditional distribution of each $\bb{Y}(t_1), \dots \bb{Y}(t_n)$, provided  $\bb{X}(t) = \bb{x}$ is not enclosed in a straight line in $\mathcal{S}$~\citep[cf.][]{ChowChau19}.
\end{enumerate}

Assumption \ref{assum:predictable} on the predictable process $\bb{X}(t)$ makes it possible to write $\bb{X}(t_i) = h(\dots, \epsilon_{t_{(i-2)}}, \epsilon_{t_{(i-1)}})$ for some measurable function $h: \R^\infty \rightarrow \chi$. It is a common assumption in related literature \citep[see, e.g.,][]{zhao2008confidence}. Assumption \ref{assum:openconvex} is an operational assumption, which in conjunction with \ref{assum:kernelmatrix}, ensures that $\mathcal{M}_{\bb{u},n}^{(p)}$  consists of a sequence of random convex functions on $\mathcal{U}_p$, for every $p \geqslant 1$; thereby forming a sequence of continuous functions. This, in turn, ascertains the existence of M-estimators $\bbhat{q}_n$ in the quality of a uniquely defined sequence of minimizers of $\mathcal{M}_{\bb{u},n}^{(p)}$. We further note that the objective function $\mathcal{M}_{\bb{u},n}^{(p)}(\bb{q})$ is strictly convex in its argument $\bb{q}$, and the same conclusion is true for $\mathcal{M}_{\bb{u}}^{(p)}(\bb{q})$ regardless of the square-integrable spatial process $u$, and for every fixed $p \in \N$. This strict convexity ensures that the true quantile $\bb{q}_0$ is a well-separated minimizer of $\mathcal{M}_{\bb{u}}^{(p)}$, that is, there exists a $\bb{q}_0 \in \mathcal{U}_p$ for which $\inf_{\bb{q}: \, \Vert \bb{q} -\bb{q}_0 \Vert \geqslant \delta }\, (\mathcal{M}_{\bb{u}}^{(p)} (\bb{q})- \mathcal{M}_{\bb{u}}^{(p)} (\bb{q}_0) ) >0$, for arbitrarily small $\delta > 0$. Assumption~\ref{assum:openconvex} also ensures that the observed points $\bb{Y}(t_i)$ are excluded from the feasible set of $\bb{q}$, to ensure well-definedness of the estimating equation introduced later in Section~\ref{sec:key-prop}.

\subsection{Estimation algorithm}\label{sec:algorithm}

The idea behind the Iteratively Re-weighted Least Squares (IRLS) type estimation as an effective means to achieve minimization in \eqref{eqn:estimator} is to replace the objective function with an equivalent version whose iterative minimization leads to the closest answer to the original problem of obtaining a quantile estimator $\hat{\bb{q}}_n(\bb{u} ,\bb{x})$. Firstly, we expand the sample criterion function  $\mathcal{M}_{\bb{u},n}$ as 
\begin{equation*}
   \frac{1}{n}\sum_{i=1}^n \frac{\left(\bb{Y}(t_i) - \bb{q}\right)\tr \bb{K}^2(t_i) \left(\bb{Y}(t_i) - \bb{q} \right)}{\Vert \bb{Y}(t_i) - \bb{q}\Vert_{\bb{K}^2(t_i)}} + \langle \bb{u}, (\bb{Y}(t_i) - \bb{q}) \rangle_{\bb{K}(t_i)},
\end{equation*}
where we use the convention that $\bb{y}/\Vert \bb{y} \Vert = 0$ if $\bb{y}= \bb{0}$. In addition, we highlight that $\bb{q}$ does not change with $i$.  Henceforth, we shall use $w_i(\bb{q}) := \Vert \bb{Y}(t_i) - \bb{q}\Vert^{-1}_{\bb{K}^2(t_i)}$ so as to avoid notational burden. Minimization of the above with respect to $\bb{q}$ yields the sequence of estimators
\begin{equation}
    \hat{\bb{q}}_n = \left[ \sum_{i=1}^n w_i(\bb{q})\bb{K}^2(t_i) \right]^{-1} \left\{ \frac{1}{2}\sum_{i=1}^n \bb{K}(t_i)\bb{u} + \sum_{i=1}^n w_i(\bb{q})\bb{K}^2(t_i) \bb{Y}(t_i)\right\}.
    \label{eqn:iteration}
\end{equation}

The above expression suggests that, given an initial estimate $\hat{\bb{q}}^{(0)}$, it is possible to use \eqref{eqn:iteration} iteratively in order to improve estimation at the next iteration. Specifically, $\bbhat{q}^{(k+1)}$ is obtained by calculating $w_i(\bbhat{q}^{(k)})$ for all $i = 1, 2, \dots n$, using the current estimate and then using the above equation to go to the next step of the iteration. Despite the matrix $ \sum_{i=1}^n w_i(\bb{q})\bb{K}(t_i)$ guaranteed to be positive definite with probability tending to one, in practice it may become numerically singular at some stage of the iterative algorithm, particularly when the number of locations $p$ is large. By adding a small positive number $\rho$ to the diagonal elements of this matrix, such a numerical complication can be easily averted. The hyperparameter $\rho$ subsequently becomes the Lagrangian coefficient associated with a new constraint for the objective function \eqref{eqn:estimator}: $\Vert \bb{q}\Vert \leqslant C$, for some fixed positive number $C$. This iterative procedure can also be viewed as a Newton-Raphson type procedure, ensuring its convergence. In particular, it is of straightforward verification that this iteration scheme reduces the value of the objective function $\Mcal_{\bb{u},n}^{(p)}(\bbhat{q})$ at the next iteration, i.e., $\Mcal_{\bb{u},n}^{(p)}(\bbhat{q}^{(k+1)}) \leqslant \Mcal_{\bb{u},n}^{(p)}(\bbhat{q}^{(k)})$ for all $k \in \N$. Combining this fact with the strict convexity of the objective function, it follows that the solution of the algorithm converges to the global minimizer of $\Mcal_{\bb{u},n}^{(p)}(\bb{q})$ since it is inherently assumed that the set of global minimizers is non-empty. This result is formulated in the next Lemma.

\begin{lemma}\label{lem:convergence}
    Let $\bb{K}(t) = \bb{K}(\bb{x}, \bb{X}(t))$ be a matrix-valued kernel function such that the maximum eigenvalue of $\bb{K}(t)$ is uniformly bounded for all $t \in [0, 1]$. Then, with an initial estimate $\bbhat{q}^{(0)}$, the sequence of improved estimators obtained by the iteration equation
    \begin{equation*}
        \bbhat{q}^{(k+1)} = \left[ \sum_{i=1}^n w_i(\bbhat{q}^{(k)})\bb{K}^2(t_i) \right]^{-1} \left\{ \frac{1}{2}\sum_{i=1}^n \bb{K}(t_i)\bb{u} + \sum_{i=1}^n w_i(\bbhat{q}^{(k)})\bb{K}^2(t_i) \bb{Y}(t_i)\right\}, \ k \geqslant 0,
    \end{equation*}
    \noindent converges as $k \rightarrow \infty$ and the ultimate estimator $\bbhat{q}^\ast$ is the solution $\bbhat{q}_n(\bb{u}, \bb{x})$ of~\eqref{eqn:estimator}. 
\end{lemma}

\subsection{Choice of the kernel and associated bandwidth}\label{sec:bandwidth-choice}

The matrix-valued kernel function $\kernel{t}$ involving the bandwidth $b_n>0$ requires suitable normalizing constants $c_n>0$. Specifically, we consider the matrix
\begin{equation*}
    \kernel{t} = c_n^{-1} \Kcal\left( \dfrac{\bb{x} - \bb{X}(t)}{b_n} \right),
\end{equation*}
given in terms of a symmetric and continuous kernel $\Kcal: \R^d \rightarrow \R^{p\times p}$, with compact support on $B^{d-1} = \{ \bb{x} \in \R^d : \Vert \bb{x}\Vert \leqslant 1\}$, and in keeping with the formulation we typically encounter for such nonparametric adjacency matrices of weights~\citep[e.g.][]{deb2017asymptotic}. Specific choices for constants $b_n, c_n >0$ must be based not only on theoretical considerations, but also informed by empirical evidence on the behavior of competing estimators in situations of repeated sampling, all considered against the backdrop of assumptions threaded in the relevant subject literature. For example, univariate nonparametric kernel estimators for a trend in time series, which have been proposed in~\cite{zhao2008confidence}, appear suitably normalized via $c_n = \sqrt{n\,b_n}$. We also assume that $\Kcal(0) = c_n\bb{I}_p$, similar to the univariate case where the kernel function at origin is taken to be equal to $1$. Given the choice of a particular form of kernel $\Kcal$, one may choose $c_n$ to be a function of the bandwidth $b_n$ such that the eigenvalues of the normalized kernel matrix $\kernel{t}$ remain bounded away from $0$ and infinity. 

Since the problem of estimating the conditional kernel tackled in here is closely related to that of conditional density estimation, existing methods with proven efficacy can be invoked \citep[see e.g.][]{Einbecketal2006}, alongside the use of graphical approaches such as an empirical scatterplot of the maximum eigenvalue of kernel matrix over different choices of $b_n$ to obtain the rate of growth of the eigenvalues. It is also possible to couple our results with the theoretical findings on the spectrum of kernel matrices for popular kernels (e.g. Gaussian kernel, Mat\'{e}rn kernel) presented in \cite{simon2021spectral}, albeit the choice of the bandwidth $b_n$ should be moderated by a cross-validation approach that verifies the theoretical conditions set out, which in this particular research are outlined in Section~\ref{Sec:Asymptotics}. A complete treatment of these and analogous conditions requires technical apparatus of which a detailed study can be found in \cite{Fanetal1996}.

\subsection{Key properties of the nonparametric quantile estimator}\label{sec:key-prop}
The results encompassing this section, summarized in Lemma~\ref{lemma:estimating-eqn} and ensuing considerations, are viewed as fundamental steps to establishing performance guarantees of the proposed nonparametric methodology for quantile estimation. Indeed, a desirable property of a quantile estimator is affine equivariance: if the samples $\bb{Y}(t)$ are changed to $\bb{A}\bb{Y}(t) + \bb{b}$ for some non-singular matrix $\bb{A}$, the estimator should also exhibit the same affine transformation. We note that the proposed quantile estimator $\bbhat{q}_n(\bb{u} ,\bb{x})$ resulting from~\eqref{eqn:estimator} is equivariant under location transformation, i.e., employing a shift to observations through $\bb{Y}(t_i) + \bb{a}$ changes the sequence of estimators accordingly to $\hat{\bb{q}}_n(\bb{u} ,\bb{x}) + \bb{a}$. Although the equivariance also holds in general with respect to a scale transformation $c\bb{Y}(t_i)$, for some scalar $c \in (0, \infty)$, verification of the orthogonal equivariance property (or more general affine equivariance property) is dependent on the incorporating kernel matrix $\bb{K}(t_i)$~\citep[see][]{chaudhuri1996geometric,Chen1996,KonenPaindaveine22}.

As highlighted in Section~\ref{sec:introduction}, quantile estimators should generally exhibit two key properties: non-crossing of the quantiles and smoothness of the estimator as a function of the data. In the following discussion, we demonstrate how upon a very minimal set of assumptions, both these properties follow in relation to $\widehat{\bb{q}}_n(\bb{u}, \bb{x})$.

For each $n \geqslant 1$, the quantile estimator $\bbhat{q}_n(\bb{u} ,\bb{x})$ is continuous with respect to observables $\bb{Y} = \left(\bb{Y}(t_1), \dots \bb{Y}(t_n)\right)$. In view of assumption~\ref{assum:openconvex}, we can without loss of generality restrict the optimization to a compact set $\mathcal{Q} \subseteq \R^p$ containing the true targeted quantile. Consequently, $\bbhat{q}_n(\bb{u} ,\bb{x}) = \bb{q}(\bb{Y}) = \argmin_{\bb{q} \in \mathcal{Q}} \mathcal{M}^{(p)}_{\bb{u},n} (\bb{q}, \bb{Y})$. Now, let $\bb{Y}_r \rightarrow \bb{Y}_0$ be a converging sequence with $r = 1, 2, \dots$, and let $\bb{q}_r = \bb{q}(\bb{Y}_r)$ for every $r \in \N$. We proceed with a subsequencing argument: for any $\{ \bb{q}_{r_k} \}_{k \in \mathbb{N}}$, the compactness of $\mathcal{Q}$ ensures that $\{ \bb{q}_{r_k} \}$ has an accumulation point $\bb{q}'$, say. Also, by definition of $\bb{q}_{r_k}$, we have that $\mathcal{M}^{(p)}_{\bb{u},n}(\bb{q}_{r_k}, \bb{Y}_{r_k}) \leqslant  \mathcal{M}^{(p)}_{\bb{u},n}(\bb{q}_0, \bb{Y}_{r_k}), \ r_k \in \N$. Since $ \mathcal{M}^{(p)}_{\bb{u},n}$ is a continuous function of $\bb{q}$ and $r_k \rightarrow \infty$, then $ \mathcal{M}_{\bb{u},n}(\bb{q}', \bb{Y}_0) \leqslant  \mathcal{M}_{\bb{u},n}(\bb{q}_0, \bb{Y}_0)$, thus entailing $\bb{q}' = \bb{q}_0$ for any subsequence, from which the continuity of the estimator ensues.

The next lemma outlines the desirable non-crossing property in any quantile estimation methodology \citep[e.g. in line with Theorem 2.1.2 of][]{chaudhuri1996geometric}. It is mainly a technical lemma aiming to justify that the M-estimator $\bbhat{q}_n(\bb{u} ,\bb{x})$, viewed as the approximate (asymptotic) solution of the convex optimization problem~\eqref{eqn:estimator}, gains the configuration of a $Z$-estimation problem for every fixed $(\bb{u} ,\bb{x})$. This is envisaged to simplify substantially the proofs for asymptotic normality. Nevertheless, because we will need to take into account the spatial dependence structure on the basis of an infinitely growing number of locations, the rate of convergence will differ from the familiar $\sqrt{n}$ for tightness mustered by predominant asymptotic theory for $Z$-estimators \citep[see][]{VanderVaart1998,Kosorok2008}.

\begin{lemma}\label{lemma:estimating-eqn}
    Under assumptions ~\ref{assum:predictable} to \ref{assum:kernelmatrix}, for every $n, p \in \N$, $\bb{x} \in \R^d$ and $\bb{u} \in B^{p-1}$, the sequence of minimizers $\widehat{\bb{q}}_n(\bb{u}, \bb{x})$ of the objective function $\Mcal_{\bb{u},n}^{(p)}$ (as in \eqref{eqn:m-u-n}) satisfies
    \begin{equation}\label{eqn:est-equation}
        \frac{1}{n} \sum_{i=1}^n \bb{K}(t_i) \left[\dfrac{\bb{K}(t_i)(\bb{Y}(t_i) - \hat{\bb{q}}_n(\bb{u} ,\bb{x}))  }{ \Vert \bb{K}(t_i)(\bb{Y}(t_i) - \hat{\bb{q}}_n(\bb{u} ,\bb{x})) \Vert } + \bb{u}\right] = \bb{0},
    \end{equation}
    for $\widehat{\bb{q}}_n(\bb{u}, \bb{x}) \notin \{ \bb{Y}(t_1), \dots, \bb{Y}(t_n) \}$. The true quantile $Q(\bb{u},\bb{x})$ in the quality of minimizer of $\Mcal_{\bb{u}}^{(p)}$ verifies the population version of \eqref{eqn:est-equation}, specifically
    \begin{equation}
        \E_{\bb{Y}(t) \mid \bb{X}(t) = \bb{x}}\left[ \dfrac{\bb{Y}(t)- Q(\bb{u} ,\bb{x}) }{\Vert \bb{Y}(t) - Q(\bb{u} ,\bb{x}) \Vert } + \bb{u}\right] = \bb{0}.
        \label{eqn:est-equation-true}
    \end{equation}
\end{lemma}

The actionable idea stemming from Lemma~\ref{lemma:estimating-eqn} -- that we may regard \eqref{eqn:est-equation} as an estimating equation for obtaining $\bbhat{q}_n(\bb{u} ,\bb{x})$ from a finite sample data with $n$ time-points and $p$ spatial points -- gives rise to an alternative Newton-Raphson type algorithm for obtaining the estimator. Notice that, for $\bb{u} = 0$, the estimating equation simply corresponds to a weighted spatial median estimate of $\bb{Y}(t_i)$ values, weighted by the covariate process $\{\bb{X}(t) : t \in \Tcal \}$ through the kernel matrix $\bb{K}(t)=\kernel{t}$. However, for $\bb{u} \neq \bb{0}$, there is an additional term creeping up in the way of an approximation bias, thereby increasing or decreasing the coordinates of $\bbhat{q}_n(\bb{u}, \bb{x})$ as appropriate in a data-driven manner. This is found pivotal to ensuring that the estimated quantile curves returned by our methodology enjoy the non-crossing property. In view of \eqref{eqn:est-equation-true}, it is clear that if a single coordinate $u_i$ of $\bb{u}$ is positive and increases in value, then the corresponding coordinate of $\bb{Y}(t) - Q(\bb{u}, \bb{x})$ must be negative and decrease further in expectation, resulting in $Q(\bb{u}, \bb{x})$ turning away from the center $\E(\bb{Y}(t))$. This justifies the non-crossing property.

\section{Asymptotic results}
\label{Sec:Asymptotics}

The nonparametric quantile estimator introduced in Section~\ref{Sec:Estimator} enjoys a number of desirable large-sample properties. This section is aimed at addressing these by allowing the number of spatial locations $p$ tend to infinity, albeit in a controlled manner relatively to $n \rightarrow \infty$, as mentioned earlier. The obtained asymptotic properties provide the theoretical justification for the distinctive features of the quantile estimation and ensuing testing methodology proposed in this paper.

We now fix some more notations in order to progress. A bounded sequence of numbers $\{x_n\}$ is throughout written as $x_n=\mathcal{O}(1)$, springing out to the notation $x_n= \mathcal{O}(y_n)$ if $x_n/y_n= \mathcal{O}(1)$. In general, if the random sequence $X_n$ is tight, we write $\|X_n\| = \mathcal{O}_p(1)$, and if $\|X_n\|/\|Y_n\| \conv{P} 0$ as $n \rightarrow \infty$, we denote this fact by $X_n= o_p(Y_n)$. In case of deterministic sequence, $x_n= o(y_n)$ denotes the convergence $\limsup_{n \rightarrow \infty} \vert x_n/y_n\vert = 0$.

Let $\Tcal_n = \{t_1, \dots, t_n\}$ denote the set of observable time-points that satisfy the smoothing condition $\max_{2\leqslant i \leqslant n} \vert t_i - t_{i-1}\vert \rightarrow 0$ as $n \rightarrow \infty$. We also require that $\Tcal_n$ is asymptotically dense in the temporal horizon $\Tcal$. In other words, for any $\tilde{t} \in \Tcal$, there exists a sequence of points $\tilde{t}_n \in \Tcal_n$ such that $\vert \tilde{t} - \tilde{t}_n\vert \rightarrow 0$.


\begin{theorem}\label{thm:consistency}
    Assume conditions~\ref{assum:predictable}-\ref{assum:kernelmatrix} hold. Suppose that the number of spatial locations $p$ is such that $p=p_n = o(n^{1/2})$ and $p_n^{-1}\log n = \mathcal{O}(1)$, as $n \rightarrow \infty$. Then, for all $\bb{x}$ lying in the compact set $\chi$, the sequence of estimators $\widehat{\bb{q}}_n(\bb{u}, \bb{x})$ of the population (theoretical) quantile $Q(\bb{u}, \bb{x})$ satisfies
    \begin{equation}\label{eqn:consistency}
        \Vert \widehat{\bb{q}}_n(\bb{u}, \bb{x}) - Q(\bb{u}, \bb{x}) \Vert = \mathcal{O}_p( p/\sqrt{n} ),
    \end{equation}
    provided a square-integrable spatial process $\bb{u}$, i.e.  $\int_{\Scal} \vert u(s)\vert^2 ds < 1$.
\end{theorem}

In~Theorem \ref{thm:consistency}, the number of spatial locations $p$ is allowed to tend to infinity but at a vanishing rate relatively to $n^{1/2}$.  It turns out that such a mild restriction on the growth rate of $p_n$ is just what it takes for ensuring the consistency of the quantile estimator paving the way for attaining a non-degenarate weak limit.

For ascertaining the asymptotic normality of the proposed quantile estimator, however, the number of spatial locations must align with the initial requirement that $\Scal_p$ becomes asymptotically dense in $\Scal$. Analogously to the temporal domain, we fix the setting where the observable spatial points $\Scal_p = \{s_1, s_2, \dots, s_p\}$ satisfy the condition $\max_{1 \leqslant j \neq k \leqslant n} d(s_j, s_k) \rightarrow 0$ as $p \rightarrow \infty$. Here, $d(s_j, s_k)$ represents the geodesic distance between the spatial coordinates $s_j$ and $s_k$. This ensures that $\Scal_p$ becomes dense in $\Scal$ with a growing number of spatial locations $p_n$ based on the geodesic metric. Also, the bandwidth $b_n$ of the kernel matrix, architected and discussed in Section~\ref{sec:bandwidth-choice} as $b_n \rightarrow 0$, must be such that $nb_n \rightarrow \infty$ as $n \rightarrow \infty$ so that the maximum eigenvalue of $\kernel{t}$ is uniformly tight.

\subsection{Assumptions}

While assumptions~\ref{assum:predictable} to \ref{assum:kernelmatrix} in Section~\ref{Sec:Estimator} lay the groundwork for this paper, the next set of assumptions~\ref{assum:finite-dim} to \ref{assum:alpha-beta} are deemed sufficient for attaining a Gaussian limiting distribution for the proposed quantile estimator.

Let $f_{t}$ denote the joint probability density function of $\bb{X}(t)$ and $\Gcal_{k}$ and $\Gcal_{k-}$ be the $\sigma$-algebra generated by $\{ \bb{Y}(t, s), \bb{X}(t): t \leqslant t_k, s \in \Scal \}$ and $\{ \bb{Y}(t, s): t < t_k, s \in \Scal \} \cup \{ \bb{X}(t): t \leqslant t_k, s \in \Scal \}$ respectively. Additionally, let $\mathcal{P}_k$ be the projection operator defined as $\mathcal{P}_k(Z) := \E(Z\mid \Gcal_{k}) - \E(Z\mid \Gcal_{k-1})$, for all $k \in \Z$. 

\begin{enumerate}[label=(B\arabic*),ref=(B\arabic*)]
     
     \item\label{assum:finite-dim} Suppose that the joint probability density function $f_{t}$ is differentiable with respect to each coordinate of $\bb{X}(t)$, for each $t \in \Tcal$. This takes care of finite-dimensional distributions of the observable process.
     
     \item\label{assum:bandwidth} The bandwidth $b_n>0$ satisfies the following boundary condition as $n \rightarrow \infty$:
     \begin{equation*}
         \max \left( b_n^{4/3}\log n , \, \frac{(\log n)^3}{nb_n^3} , \frac{ (\log n)^2}{b_n^{4/3}} \frac{\Xi_n}{n^2}\right) = o(1),
    \end{equation*}
    with the cumulative effect of the first error $\bb{\xi}(0)$ in predicting the covariate process $\bb{X}(n)$ defined as 
    $\Xi_n  := n\Theta_{2n}^2 + \alpha_n$ \citep[see][]{zhao2008confidence} where
    \begin{align*}
        \Theta_n & = \sum_{i=1}^n \sup_{x\in \chi} \Vert \mathcal{P}_0(f_t(\bb{x} \mid \Gcal_i)) \Vert + \sup_{x\in \chi} \Vert \mathcal{P}_0(f_t\dash(\bb{x} \mid \Gcal_i)) \Vert,\\
        \alpha_n & = \lim_{m \rightarrow \infty} \sum_{k = n}^{m} \left( \Theta_{n+k} - \Theta_k \right)^2.
    \end{align*}    
    
    \item\label{assum:kernelbound} There exists a constant $c > 1$ such that
    \begin{equation*}
        \dfrac{1}{c} < \sup_{t \in \Tcal} \dfrac{\lambda_1(\bb{K}(t))}{\lambda_p(\bb{K}(t))} < c,
    \end{equation*}
    where $\bb{K}(t)= \kernel{t}$ is the $p\times p$ kernel matrix as before and $\lambda_1(\bb{A})$ and $\lambda_p(\bb{A})$ respectively denote the largest and the smallest eigenvalues of the matrix $\bb{A}$.
    
    \item\label{assum:alpha-beta} The number of spatial points $p_n$ is such that $p_nb_n \rightarrow 0$, $p_n^{9/8}b_n \rightarrow \infty$ and $p_n^{17/8}/n \rightarrow 0$, as $n \rightarrow \infty$. 
    
\end{enumerate}

Assumption~\ref{assum:finite-dim} is natural and indeed a customary assumption that ensures smoothness of the probability density functions, allowing to employ the standard techniques in asymptotic analysis. Assumption~\ref{assum:bandwidth} is a further restriction to the growth of the bandwidth $b_n$ of the kernel matrix, which exerts influence on the strength of dependence of the process. From~\ref{assum:bandwidth}, it results that $b_n^{-1} (n^{-1/3} \log n) = o(1)$, thus imposing a lower bound on $b_n$, and $b_n (\log n)^{3/4} = o(1)$, which provides an upper bound for $b_n$. A popular  choice for $b_n>0$ is that of the type $c_0 n^{-\beta}$ for some (fixed) constants $c_0>0$ and $\beta > 0$. If it is used in tandem with the kernel proposed in Section~\ref{sec:bandwidth-choice}, we find that $b_n = n^{-1/5}$, a sequence that has been singled out  in~\citet{zhao2008confidence} for producing satisfactory results.

Next, we note that the third term embedded in the triplet in \ref{assum:bandwidth} implies  $\Xi_n/n^{2} = o(1)$ due to the earlier condition $b_n^{-1} (n^{-1/3} \log n) = o(1)$ in conjunction with $nb_n \rightarrow \infty$. Because each summand of the partial $\Theta_n$ gives the contribution of $\epsilon_0$ (as in~\ref{assum:predictable}) to the associated predictor $\bb{X}(t_{i+1})$, this term plays a relevant role in moderating the dependence of $\bb{X}(t)$. In particular, the random vectors $\bb{X}(t_{i})$ must satisfy a summability property designed to maintain the integrity of the predictor process $\bb{X}(t)$ by way of ensuring that the covariance structure of the spatio-temporal process is well-defined and can always be estimated. On the whole, this assumption has a number of parallels with published works on inference for spatio-temporal processes \citep[see e.g.][]{zhao2008confidence, zhou2009local, deb2021nonparametric}). According to \cite{zhao2008confidence}, a condition such as this corrals virtually the full spectrum for the strength of dependence, from the boundary case of the predictor process $\bb{X}(t)$ exhibiting only short-range dependence (SRD, determined by $\lim_{n \rightarrow \infty} \Xi_n < \infty$) to a weakening long-range dependence (LRD). 

Assumption~\ref{assum:kernelbound} on the other hand is required in order to control random oscillations of the kernel matrix. In keeping with the random matrix theory, Wigner's law asserts that the eigenvalues of the $p \times p$ covariance matrix $\bb{\Sigma}(\bb{X}(t))$ are of order $\sqrt{p}$~\citep{vershynin2010introduction} for sufficiently large $p$, hence monotonically increasing with $n$ and $p$. This means that for any fixed $t \in \Tcal$, the ratio of the largest and the smallest eigenvalue of $\bb{K}(t)$ remains bounded for sufficiently large $p$. For the first finitely many terms, boundedness of this ratio is ensured by tightness and positive definiteness of the kernel matrix $\bb{K}(t)$. Assumption~\ref{assum:kernelbound} is a stronger version of this statement that requires uniform boundedness over all choices of $t \in \Tcal$. This assumption guarantees that there exists a suitable kernel matrix such that the eigenvalues of the covariance matrices of the kernel-weighted observables $\bb{Y}(t)$ remain bounded. Finally, assumption~\ref{assum:alpha-beta} in conjunction with the bounds on bandwidth $b_n$ that ensue from \ref{assum:bandwidth} gives that $(\log n)^{-2/3} \ll  p_n \ll  \left( n/\log n\right)^{1/3}$, for large enough $n$. This speaks to the ultra-dimension element in our methods.

\subsection{Asymptotic distribution}

In this section, we establish the Gaussian limiting distribution attained by the proposed quantile estimator.  The consistency of the proposed quantile estimator expounded in Theorem~\ref{thm:consistency} and accompanying text herald the rate at which the number of locations $p$ can grow in relation to the number of time-points $n$ so as to make this nonparametric quantile estimation feasible. 


Continuing along the lines of \eqref{eqn:m-u-n} and  adding to the notation introduced as part of Lemma~\ref{lemma:estimating-eqn}, we define
\begin{equation*}
    \nabla \Mcal_{\bb{u},n}^{(p)} (\bb{q}) := -\dfrac{1}{n}\sum_{i=1}^n \bb{K}(t_i) \left[ \dfrac{\bb{K}(t_i) (\bb{Y}(t_i) - \bb{q})}{\Vert \bb{Y}(t_i) - \bb{q} \Vert_{\bb{K}(t_i)}} + \bb{u} \right]\label{eqn:m-u-n-prime}.
\end{equation*}
\noindent Let $\nabla^2 \Mcal_{\bb{u},n}^{(p)} (\bb{q}_0)$ be the matrix whose $i$-th row is comprised of the gradient of $\nabla \Mcal_{\bb{u},n}^{(p)} (\bb{q})$ with respect to the $i$-th coordinate of $\bb{q}$. Also, let us denote by $\nabla^3 \Mcal_{\bb{u},n}^{(p)}(\bb{q})$ the third order tensor whose $(i,j,k)$-th entry is the third order derivative of $\Mcal_{\bb{u},n}^{(p)}(\bb{q})$ with respect to the $i,j$ and $k$-th coordinate of $\bb{q}$. We have the following proposition.

\begin{proposition}\label{prop:P1P3}
Under assumptions~\ref{assum:predictable}-\ref{assum:kernelmatrix} and~\ref{assum:finite-dim}-\ref{assum:alpha-beta}, the following hold.
\begin{enumerate}[label=(P\arabic*),ref=(P\arabic*)]
    
    \item\label{prop:first-div-dist} There exists a $p$-variate Gaussian random variable $\bb{Z}(\bb{q}_0)$ with mean parameter $\eta_t(\bb{q}_0)$ and dispersion matrix $\Omega_t(\bb{q}_0)$, such that 
    \begin{equation*}
        \sup_{\bb{u} \in \mathcal{B}_p}\sup_{h \in \Hcal} \left\Vert \E\left[ h\left( (pb_n)^{-1} \nabla \Mcal_{\bb{u},n}^{(p)} (\bb{q}_0) \right) \right] - \E\left[ h(\bb{Z}(\bb{q}_0) \right] \right\Vert = \mathcal{O}_p(p^2/n),
    \end{equation*}
    where $\mathcal{B}_p$ is a sequence of compact subsets of $B^{p-1}$, $\Hcal$ is the set of bounded thrice differentiable functions from $\R^{p}$ to $\R$ with a bounded derivative up to the third order. 
        
    \item\label{prop:second-div-mat} There exists a collection of positive definite matrices $\{ \Psi_t(\bb{q}_0) \}$ for all possible $\bb{q}_0 \in \mathcal{Q}_p$ as a minimizer of \eqref{eqn:parameter} with respect to the spatial points $\Scal_p$, such that 
    \begin{equation*}
        \sup_{\bb{q}_0 \in \mathcal{Q}_p} \left\Vert \nabla^2 \Mcal_{\bb{u},n}^{(p)}(\bb{q}_0)-\Psi_t(\bb{q}_0) \right\Vert \xrightarrow{P} 0.
    \end{equation*}
    \item\label{prop:third-div-bound} There exists $M > 0$ (not depending on $n$) such that 
    \begin{equation*}
       \sup_{\bb{q} \in \mathcal{Q}_p} \left\Vert \nabla^{3} (\Mcal_{\bb{u},n}^{(p)})(\bb{q}) \right\Vert \leqslant M.
    \end{equation*}
\end{enumerate}
    
\end{proposition}

The three parts~\ref{prop:first-div-dist}-\ref{prop:third-div-bound} embodying Proposition~\ref{prop:P1P3}, whose proof is deferred to the supplementary material,  mirror each of the usual necessary steps to demonstrating the asymptotic normality of $M$-estimators~\citep[][]{huber2011robust}. The asymptotic development for the proposed estimator of the conditional quantile of $\bb{Y}(t)$, given $\bb{X}(t) = \bb{x}$ for any fixed $t \in \Tcal$, stays within this realm. Although general in their principle, the innovation in Proposition~\ref{prop:first-div-dist} is that it amplifies complementary works on quantile regression by enabling the increasing dimension $p=p_n \rightarrow \infty$, as $n \rightarrow \infty$. The simpler i.i.d. version analogous to this proposition can be traced back to \cite{portnoy1986clt}, and contemporary related results accommodating a changing $p_n$ can be found in~\cite{das2020clt}. Our infinite-dimensional framework shall build on a combination of (pre-asymptotic) moment-type results approached by a functional central limit theorem for martingale differences~\citep[cf.][]{belloni2018high}.

In order to lessen the notational burden throughout the remainder of this paper, we collect the relevant building blocks in the following: 
\begin{align}
    \bb{\eta}(t_i) &= \E\left[ \bb{K}(t_i)(\bb{Y}(t_i) - \bb{q}_0) \mid \Gcal_{t_i-} \right] = \bb{K}(t_i)\left( \bb{\mu}(\bb{X}(t_i)) - \bb{q}_0 \right),\label{eqn:notation-eta}\\
    \bb{V}_1(t_i) &= \var\left[ \bb{K}(t_i)(\bb{Y}(t_i) - \bb{q}_0) \mid \Gcal_{t_i-} \right] = \bb{K}(t_i) \bb{\Sigma}(\bb{X}(t_i)) \bb{K}(t_i),\label{eqn:notation-v1}\\
    \bb{V}_2(t_i) &= \left( \dfrac{1}{\Vert \bb{\eta}(t_i) \Vert}\bb{I}_p - \dfrac{\bb{\eta}(t_i)\bb{\eta}(t_i)\tr}{\Vert \bb{\eta}(t_i)\Vert^3} \right),\label{eqn:notation-v2}
\end{align}
where $\bb{I}_p$ denotes the identity matrix of order $p$. Additionally, the expectation and covariance pertaining to the finite-dimensional distributions of the quantile process are encapsulated in:
\begin{align}
    \bb{\eta}_t(\bb{q}_0) := & \dfrac{1}{npb_n}\sum_{i=1}^n \bb{K}(t_i) \left[ \dfrac{\bb{\mu}(\bb{X}(t)) - \bb{q}_0}{\Vert \bb{\mu}(\bb{X}(t)) - \bb{q}_0 \Vert} - \dfrac{\bb{\eta}(t_i)}{\Vert \bb{\eta}(t_i) \Vert} \right], \label{eqn:keydef-eta}\\
    \bb{\Omega}_t(\bb{q}_0) := & \dfrac{1}{n^2p^2b_n^2} \sum_{i=1}^n \bb{K}(t_i) \bb{V}_2(t_i)\bb{V}_1(t_i)\bb{V}_2(t_i)\tr \bb{K}(t_i), \label{eqn:keydef-omega}\\
    \bb{\Psi}_t(\bb{q}_0) := & \dfrac{1}{n}\sum_{i=1}^n \bb{K}(t_i) \bb{V}_2(t_i) \bb{K}(t_i)\label{eqn:keydef-psi}.
\end{align}

We arrive at the main result in this paper, stated in the following theorem.

\begin{theorem}\label{thm:normal-dist}
    Consider $\bb{q}_0: = Q(\bb{u}, \bb{x})$. Under the assumptions~\ref{assum:finite-dim}-\ref{assum:alpha-beta} and in the setup of Theorem~\ref{thm:consistency}, suppose  $p^{17/8}/n \rightarrow 0$, as $n \rightarrow \infty$. Then, with $\Hcal$ from~\ref{prop:first-div-dist}, it holds that
    \begin{equation*}
        \sup_{\bb{u} \in \mathcal{B}_p} \sup_{h \in \Hcal} \left\Vert \E\Bigl[ h\left((pb_n)^{-1} (\widehat{\bb{q}}_n(\bb{u},\bb{x}) - \bb{q}_0 \right)\Bigr] - \E\Bigl[ h\left( -\bb{\Psi}_t(\bb{q}_0)^{-1}\bb{Z}(\bb{q}_0) \right) \Bigr]\right\Vert \xrightarrow{P} 0, 
    \end{equation*}
    where $\mathcal{B}_p$ is any sequence of compact subsets of $B^{p-1}$, $\bb{Z}(\bb{q}_0)$ is a $p$-variate Gaussian random vector with mean vector $\bb{\eta}_t(\bb{q}_0)$ and dispersion matrix $\bb{\Omega}_t(\bb{q}_0)$ furnished in \eqref{eqn:keydef-eta} and \eqref{eqn:keydef-omega}.
\end{theorem}

At this point, we tactically turn focus to sketching a proof for Proposition~\ref{prop:P1P3} since this forms the basis for the proof of Theorem~\ref{thm:normal-dist}. For the first part in reference, we start by noting that the challenge in obtaining the asymptotic distribution of $(pb_n)^{-1} \nabla\Mcal_{\bb{u},n}^{(p)}(\bb{q}_0)$ stems from the non-linearity of the normalized term
$
   \bigl(  \bb{K}(t_i)(\bb{Y}(t_i) - \bb{q}_0)\bigr) / \Vert \bb{K}(t_i)(\bb{Y}(t_i) - \bb{q}_0) \Vert
$
with respect to the random variables $\bb{Y}(t_i)$. A standard trick to thwart this issue is to employ a suitable Taylor's expansion for obtaining a sufficiently sharp approximation to a linear functional of  $\bb{Y}(t_i)$ or $\bb{\xi}(t_i)$. Indeed, the expressions \eqref{eqn:notation-eta}-\eqref{eqn:notation-v2} have already been carved out in such a way to prepare for such an expansion: writing $\bb{K}(t_i)(\bb{Y}(t_i) - \bb{q}_0) = \bb{\eta}(t_i) + \bb{V}_1^{1/2}(t_i)\bb{\xi}(t_i)$. Application of Taylor's expansion on the errors $\xi_{tj}$ thus precipitates the development
\begin{equation*}
    \dfrac{\bb{K}(t_i)(\bb{Y}(t_i) - \bb{q}_0)}{\Vert \bb{K}(t_i)(\bb{Y}(t_i) - \bb{q}_0) \Vert} = \dfrac{\bb{\eta}(t_i)}{\Vert \bb{\eta}(t_i) \Vert} + \bb{V}_2(t_i)\bb{V}_1^{1/2}(t_i) \bb{\xi}(t_i) + \mathcal{O}\left(\dfrac{\Vert \bb{V}_1^{1/2}(t_i)\Vert}{\Vert \bb{\eta}(t_i)\Vert^3} \Vert \bb{\xi}(t_i) \Vert^2\right),
\end{equation*}
i.e.,
\begin{multline*}
    \nabla \Mcal_{\bb{u},n}^{(p)}(\bb{q}_0)
    = \dfrac{1}{n} \sum_{i=1}^n \bb{K}(t_i) \left( \dfrac{\bb{\eta}(t_i)}{\Vert \bb{\eta}(t_i) \Vert} + \bb{u} \right) + \dfrac{1}{n} \sum_{i=1}^n \bb{K}(t_i) \bb{V}_2(t_i)\bb{V}_1^{1/2}(t_i) \bb{\xi}(t_i)\\
    + \dfrac{1}{n}\sum_{i=1}^n \mathcal{O}(\Vert \bb{K}(t_i) \Vert \Vert \bb{\eta}(t_i)\Vert^{-3} \Vert \bb{V}_1^{1/2}(t_i) \Vert \Vert \bb{\xi}(t_i) \Vert^2).
\end{multline*}

The proof now unfolds in three steps. Owing to Lemma~\ref{lemma:estimating-eqn}, the first term can be shown to converge to a constant which constitutes the bias. Next, application of a bespoke central limit theorem for the multidimensional martingale provided by~\citet{belloni2018high}, ensures that the asymptotic weak limit to the second term is a Gaussian random variable with mean zero. Finally, by assumption~\ref{assum:alpha-beta}, the third term vanishes to zero at the rate $\mathcal{O}(\max\{ p^{-9/4}b_n^{-2}, p^{-3/2}b_n^{-1} \})$.  Further details concerning this proof for Proposition~\ref{prop:first-div-dist} are expounded in Appendix A1 as part of the supplementary material accompanying this paper.

As for the proof of Proposition~\ref{prop:second-div-mat}, it is here sketched as follows (the proof in its entirety is deferred to Appendix A2). The second order derivative of the objective function $\Mcal_{\bb{u},n}^{(p)}(\bb{q})$ is calculated in the form:
\begin{equation*}
    \nabla^2 \Mcal_{\bb{u},n}^{(p)}(\bb{q}) 
    = \dfrac{1}{n}\sum_{i=1}^n \bb{K}(t_i) \biggl[ \dfrac{1}{\left\Vert  \bb{Y}(t_i) - \bb{q} \right\Vert_{\bb{K}(t_i)} }\bb{I}_p 
    - \dfrac{\bb{K}(t_i)(\bb{Y}(t_i) - \bb{q})(\bb{Y}(t_i) - \bb{q})\tr \bb{K}(t_i)}{\left\Vert  \bb{Y}(t_i) - \bb{q} \right\Vert^3_{\bb{K}(t_i)}} \biggr] \bb{K}(t_i).
\end{equation*}

After splitting the above expression into a sum of martingale difference terms and a remainder term, the former sum, suitably normalized, can be shown to converge to zero via application of a martingale central limit theorem~\citep[see][]{belloni2018high}. The dominant component in the remainder term  boils down to $\bb{\Psi}_t(\bb{q}_0)$ with everything else being asymptotically negligible. This is achieved through a dedicated Taylor's expansion.

Finally, regarding Proposition~\ref{prop:third-div-bound}, we find it convenient and useful to prove the stronger statement that $\nabla^2 \Mcal_{\bb{u},n}^{(p)}(\bb{q})$ is Lipschitz with a Lipschitz constant $M$ independent of $n$ and $p$.  Starting with assumption~\ref{assum:openconvex}, we note that since $\mathcal{Q}_p$ is a compact subset of $\mathcal{U}_p$, $\Vert \bb{Y}(t_i) - \bb{q} \Vert_{\bb{K}(t_i)} \geqslant \lambda_p(\bb{K}(t_i)) \delta$ for all $\bb{q} \in \mathcal{Q}_p$. Since the kernel matrix is positive definite and has bounded entries,  it readily follows that $\nabla^2 \Mcal_{\bb{u},n}^{(p)}(\bb{q})$ is a Lipschitz function of $\bb{q}$ with a Lipschitz constant $\bigl(\delta^{-1} \sup_{t_i \in \Tcal_p} \lambda_1(\bb{K}(t_i))/\lambda_p(\bb{K}(t_i))\bigr)^{-3}$. By assumption~\ref{assum:kernelbound}, this quantity remains bounded.

\section{Statistical inference: interval estimation and hypotheses testing}\label{sec:ht}

While the estimation of quantile curves yielding from \eqref{eqn:estimator} possesses several desirable qualities, including the consistency and asymptotic normality, the underpinning convergence is pointwise in the temporal domain, that is, it is obtained for each value of the covariate process $\{\bb{X}(t) = \bb{x}\}$. From the practical stance, an interesting approach resides in obtaining simultaneous confidence band. Inevitably, this will require extending the asymptotic properties of the estimator to the entire temporal domain, uniformly on the index range of $\bb{X}(\cdot)$. The means to achieve this has a number of parallels with the techniques exploited in \cite{zhao2008confidence}.

To this end, we shall consider a set of values for the covariates, sufficiently apart in order that at every time-point $t$, the kernel estimator is affected by at most one of the values in that set. This is to ensure that no correlation between estimates are at play for sufficiently distant time-points. Upon the already obtained asymptotic normality results, this consideration opens way to making uniform statistical inference possible: asymptotic behavior of the proposed nonparametric estimator's sequence across multiple time-points, uniformly, is now within reach. The ultimate aim is to devise meaningful asymptotic results pertaining to partial maxima of this sequence of estimators by drawing on extreme value theory.

We define $\chi_n = \{ \bb{X}(t): t \in \Tcal_n \} \subseteq \chi$ as the range of the covariate process $\{\bb{X}(t)\}$. We assume that, for any $\bb{x}, \bb{x}\dash \in \chi_n$, $\Vert \bb{x} - \bb{x}\dash\Vert > 2b_n\rightarrow 0$ thus implying that $\chi_n$ becomes dense within $\chi$ as $n$ tends to infinity. The cardinality of  $\chi_n$ is encapsulated in $m_n$.

\subsection{Simultaneous confidence bounds} 

For the sake of simplicity, the results comprising in this section refrain to $\bb{u}$-quantiles with $\bb{u} = (2\tau - 1)\bb{1}_p/\sqrt{p}$ for some $\tau \in (0,1)$. Henceforth, we denote this as $\bb{u}_{\tau}$. This framework essentially renders quantiles that spread evenly among the considered spatial locations. We shall denote the corresponding quantile estimator by $\bbhat{q}_n(\tau, \bb{x})$ (or sometimes simply by $\bbhat{q}(\tau, \bb{x})$) whereas the true targeted quantile is denoted by $Q(\tau, \bb{x})$. We emphasize that this slight restriction to the framework is imposed for convenience, but all results derived in this section can be adapted to incorporate the more general multivariate case as before, in a relatively straightforward manner.

By construction, it follows from the definition of $\chi_n$ that, for any two points $\bb{x}_{i_1}, \bb{x}_{i_2} \in \chi_n$, $i_1 \neq i_2$, and every $\delta>0$, the exists $n_0 \in \mathbb{N}$ such that $\norm{\bb{x}_{i_1} - \bb{x}_{i_2}} < 2\delta$, $i_1, \, i_2 > n_0$. Hence, for every $t \in \Tcal_n$, $\Vert \bb{X}(t) - \bb{x}_{j}\Vert < \delta$, $j=i_1, i_2$, which results from the triangle inequality. Since $\Kcal$ has bounded support in the unit hypersphere $B^{d-1}$, it thus follows that the cross products of the weights intervening in the objective function resolve to zero, i.e.
\begin{equation}\label{eqn:nullcrossproduct}
    \bb{K}(\bb{x}_{i_1}, \bb{X}(t)) \bb{K}(\bb{x}_{i_2}, \bb{X}(t)) = c_n^{-2} \Kcal\left( \dfrac{\bb{X}(t)-\bb{x}_{i_1}}{b_n} \right) \Kcal\left( \dfrac{\bb{X}(t)-\bb{x}_{i_2}}{b_n} \right) = 0
\end{equation}
(the reader may wish to recall the discussion on $c_n$ as part of Section~\ref{sec:bandwidth-choice}). Equation \eqref{eqn:nullcrossproduct} lays bare that fixing two quantile levels $\tau$ and $\tau\dash$ ($0 < \tau, \tau\dash < 1$), so that objective function if optimized for $\bbhat{q}(\tau, \bb{x}_{i_1})$ contains the terms with weights $\bb{K}(\bb{x}_{i_1}, \bb{X}(t))$ for some $t \in \Tcal_n$, whereas for the sequence of estimators $\bbhat{q}(\tau\dash, \bb{x}_{i_2})$ these weights amount to $\bb{K}(\bb{x}_{i_2}, \bb{X}(t))$, results in the two objective functions (both for $\bbhat{q}(\tau, \bb{x}_{i_1})$ and $\bbhat{q}(\tau\dash, \bb{x}_{i_2})$) being uncorrelated, thus implying that the estimators $\bbhat{q}(\tau, \bb{x}_{i_1})$ and $\bbhat{q}(\tau\dash, \bb{x}_{i_2})$ are also uncorrelated.

Therefore, by taking a multitude of points on the set $\chi_n = \{ \bb{x}_{i_1}, \bb{x}_{i_2}, \dots \bb{x}_{i_{m_n}} \}$, we obtain from Theorem~\ref{thm:normal-dist} that the estimated process $\{ \hat{q}_n(\tau, \bb{x})(s) : s \in \Scal \}$ becomes asymptotically a Gaussian process on both temporal and spatial domains, for every $\bb{x} \in \chi_n$. More concretely, on the finite set of spatial points $S_d = \{s_1, s_2, \dots s_d\}$ and with unit vector $\bb{l} \in \R^d$, we have for every $\bb{x} \in \chi_n$, as $n \rightarrow \infty$, that
\begin{equation}
   R_n(\bb{x}, \bb{l}, S_d) =(p b_n )^{-1} \dfrac{\bb{l}\tr (\hat{q}_n(\tau, \bb{x})\vert_{S_d} - Q(\tau, \bb{x})\vert_{S_d} - \bb{\mu}_q(\bb{x})\vert_{S_d} ) }{ \sqrt{\bb{l}\tr \bb{\Sigma}_q(\bb{x})\vert_{S_d} \bb{l} }} \xrightarrow{d} \normdist(0, 1).
\end{equation}

We highlight that $\hat{q}_n(\tau, \bb{x})\vert_{S_d}$ and $Q(\tau, \bb{x})\vert_{S_d}$ respectively denote spatial process $\hat{q}_n(\tau, \bb{x})$ and $Q(\tau, \bb{x})$ restricted to the spatial locations in $S_d$. Letting $q_0 = Q(\tau, \bb{x})$, then $\bb{\mu}_q(\bb{x})\vert_{S_d}$ and $\bb{\Sigma}_q(\bb{x})\vert_{S_d}$ respectively denote mean and covariance structure, evaluated at the spatial locations in the set $S_d$, of the Gaussian process $\{ G(\bb{x}, s) : \bb{x} \in \chi, s \in \Scal \}$ characterized by
\begin{equation}
	\begin{split}
	\E\left[ G(\bb{x}, s) \right] & = \bb{\Psi}_{\bb{X}(t) = \bb{x}}^{-1}({q}_0) \bb{\eta}_{\bb{X}(t) = \bb{x}}({q}_0) \\
	\text{Cov}\left( G(\bb{x}, s), G(\bb{x}', s) \right) & = \begin{cases}
        \bb{\Psi}_{\bb{X}(t) = \bb{x}}^{-1}({q}_0) \bb{\Omega}_{\bb{X}(t) = \bb{x}}({q}_0) \bb{\Psi}_{\bb{X}(t) = \bb{x}}^{-1} ({q}_0), & \bb{x} = \bb{x}'; \\
        0, & \bb{x} \neq \bb{x}'.
    \end{cases}
	\end{split}
 \label{eqn:gaussian-process}
\end{equation}
where the terms $\bb{\eta}({q}_0), \bb{\Omega}({q}_0)$ and $\bb{\Psi}({q}_0)$ are evaluated at $\bb{X}(t) = \bb{x}$ as defined in~\eqref{eqn:keydef-eta}-\eqref{eqn:keydef-psi}. 

We now proceed to obtaining the asymptotic behavior of the supremum of $G(\bb{x}, s)$ over the temporal index through $\sup_{\bb{x} \in \chi_n} \hat{q}_n(\tau, \bb{x})(s)$ for any $s \in \Scal$.  From the case in point that $\hat{q}_n(\tau, \bb{x}_{i_1})$ and $\hat{q}_n(\tau\dash, \bb{x}_{i_2})$ are asymptotically uncorrelated, the next step is to apply 
the extreme value theorem upon the  univariate martingale difference sequence $R_n(\bb{x}_{i_k}, \bb{l}, S_d)$, $k = 0, 1, \dots,m_{n} - 1$. Owing to the rich theory of extreme values, there exists normalizing functions $ C_t$ specific to the folded normal distribution,
\begin{equation}
    C_t(z) := \frac{1}{\sqrt{2\log t}}  \left( z -  \frac{\log \log t  + \log(4 \pi)}{2} \right)
    \label{eqn:NormingBn}
\end{equation}
\citep[see][p.72]{Resnick1987}, such that the weak convergence to the Gumbel distribution function is attained in the following sense: for all $z \in \mathbb{R}$, 
\begin{equation}
    \ \lim_{n\rightarrow \infty} \prob\left( \sup_{\bb{x} \in \chi_{m_{n}}} \sup_{\norm{\bb{l}} = 1} \vert R_n(\bb{x}, \bb{l}, S_d) \vert \leqslant C_{m_{n}}(z + 2 \log m_{n} ) \right)  = \exp\{  -e^{-z}\}.
    \label{eqn:conf-1}
\end{equation}

Next, we consider the identification
\begin{equation*}
    \sup_{\norm{\bb{l}} = 1} \dfrac{\bb{l}\tr \bb{A}\bb{l}}{\bb{l}\tr \bb{B}\bb{l}} =  \lambda_1 \left(\bb{B}^{-1/2}\bb{A}\bb{B}^{-1/2} \right),
\end{equation*}
where $\bb{A}$ and $\bb{B}$ are positive definite matrices. In conjunction with the limiting distribution statement in~\eqref{eqn:conf-1}, we subsequently find that
\begin{multline}
    \lim_{n\rightarrow \infty} \prob\biggl( \sup_{\bb{x} \in \chi_{m_{n}}} \dfrac{1}{pb_n}\lambda_1\left[\bb{\Sigma}_q(\bb{x})^{-1/2}\vert_{S_d} \bb{e}(\bb{x})\vert_{S_d}\, \bb{e}(\bb{x})\vert_{S_d}\tr \bb{\Sigma}_q(\bb{x})^{-1/2}\vert_{S_d} \right] \\ < C_{m_n}\left(z + 2 \log m_{n}\right)\biggr) = \exp\{  -e^{-z}\},
    \label{eqn:conf-2}
\end{multline}
by taking supremum over all possible unit vectors $\bb{l}$ and whereby $\bb{e}(\bb{x})\vert_{S_d} = (\hat{q}_n(\tau, \bb{x})\vert_{S_d} - Q(\tau, \bb{x})\vert_{S_d} - \bb{\mu}_q(\bb{x})\vert_{S_d} )$. Hence, under the assumptions~\ref{assum:predictable}-\ref{assum:kernelmatrix} and~\ref{assum:finite-dim}-\ref{assum:alpha-beta}, with a judicious choice of $\chi_{m_n}$, i.e., so long as $\log m_n = o \left(  (p b_n )^{-2}\right)$, the sequence of estimators $\hat{q}_n(\tau, \bb{x})$ can be guaranteed to fulfill \eqref{eqn:conf-1} and~\eqref{eqn:conf-2} on a finite set of spatial locations $S_d = \{ s_1, s_2, \dots s_d \}$ from the ``larger'' set $\Scal$. This result, which resonates with that in Theorem 2 of~\cite{zhao2008confidence}, serves as basis to introducing simultaneous confidence bands for the $\tau$-quantile curves $Q(\tau, \bb{x})$ ascribed to multiple values $\bb{x} \in \chi_n$. Because $\chi_n$ becomes increasingly dense in $\chi$ as $n$ tends to $\infty$, an asymptotic confidence band for the quantile curve $\{Q(\tau, \bb{x}): \bb{x} \in \chi\}$ arises from \eqref{eqn:conf-2}. 

A contrast-like linear combination of the quantile curves across different locations stems from \eqref{eqn:conf-1} as follows: with a proper specification of the set $S_d$ and vector $\bb{l} \in \R^d$ so that $\bb{l}\tr Q(\tau, \bb{x})$ becomes the desired contrast-like quantile curve, we find that the asymptotic simultaneous $100(1-\alpha^*)\%$-confidence bounds for $\bb{l}\tr Q(\tau, \bb{x})\vert_{S_d}$ consist of 
\begin{equation*}
    \bb{l}\tr \hat{q}_n(\tau, \bb{x})\vert_{S_d} - \bb{l}\tr\widehat{\bb{\mu}}_q(\bb{x})\vert_{S_d} 
    \pm \left[ C_{m_n}\left(-\log\left(-\log(1-\alpha^*\right)) + 2\log m_n \right) \right] (pb_n) \left( \bb{l}\tr \widehat{\bb{\Sigma}}_q(\bb{x})\vert_{S_d} \bb{l} \right)^{1/2},
\end{equation*}
with embedded consistent estimators $\widehat{\bb{\mu}}_q(\bb{x})$ and $\widehat{\bb{\Sigma}}_q(\bb{x})$ for the bias term $\bb{\mu}_q(\bb{x})$ and covariance $\bb{\Sigma}_q(\bb{x})$, respectively. We note that the symmetry in the confidence interval is justified by the same limiting extreme value distribution of $\inf_{\bb{x}}R_n(\bb{x},\bb{l},S_d)$ and $-\sup_{\bb{x}}\abs{R_n(\bb{x},\bb{l},S_d)}$, a remark that will prove useful to constructing one-sided hypothesis testing procedures. 


\subsection{Two tests}\label{subsec:tests}

The statement in \eqref{eqn:conf-2} can be adapted to the purpose of hypothesis testing aimed at statistically validating structural assumptions on the quantile curves across all locations simultaneously. Under the null hypothesis about the structural restriction of the quantile curve $Q(\tau, \bb{x})$ at some specific quantile $\tau \in [0, 1]$, we can calculate the observed value of the test statistic 
\begin{equation}\label{eqn:errors}
    \bb{e}_0(\bb{x})\vert_{\Scal_p} = (\hat{q}_n(\tau, \bb{x})\vert_{\Scal_p} - \hat{q}_0(\tau, \bb{x})\vert_{\Scal_p} - \bbhat{\mu}_q(\bb{x})\vert_{\Scal_p} )
\end{equation}
where $\hat{q}_0(\tau, \bb{x})$ stands as the estimate of $Q(\tau, \bb{x})$ under the postulated null hypothesis $H_0$ regarding a stable regime, and $\bbhat{\mu}_q(\bb{x})$ is the estimate of the mean of the Gaussian process introduced in~\eqref{eqn:gaussian-process} under the null hypothesis. This can be done by replacing $\bb{\mu}(\bb{X}(t))$ and $\bb{q}_0$ by the estimated median $\bbhat{q}_n(\bb{0}, \bb{X}(t))$ and the estimated quantile $\bbhat{q}_n(\bb{u}, \bb{X}(t))$ under $H_0$ in the corresponding terms. Then, the null hypothesis is rejected if \eqref{eqn:errors} exhibits large oscillation, i.e., at the nominal level of the test $\alpha^*$, $H_0$ is rejected if
\begin{equation*}
    \dfrac{1}{pb_n}\sup_{\bb{x} \in \chi_n} \lambda_1\left( \widehat{\bb{\Sigma}}_q(\bb{x})^{-1/2}\vert_{\Scal_p} \bb{e}_0(\bb{x})\vert_{\Scal_p}\bb{e}_0(\bb{x})\vert_{\Scal_p}\tr \widehat{\bb{\Sigma}}_q(\bb{x})^{-1/2}\vert_{\Scal_p} \right)
\end{equation*}
exceeds the critical value $c_{\alpha^*}=C_{m_n}\left(-\log\left(-\log(1-\alpha^*)\right) + 2 \log m_n \right)$. 


A structural assumption that is often put to testing is that of homogeneity of the covariate effect across the spatial domain, which, if true, will automatically narrow down the spatio-temporal structure of the data to a temporal one. This is statistically parlayed in the null hypothesis $H_0^{(1)}:\, Q(\tau, \bb{x}) = \beta(\bb{x}) \bb{u}_{\tau}$, for all $\tau \in [0, 1]$. Under $H_0^{(1)}$, the estimator $\widehat{\beta}(\bb{x})$ arises from~\eqref{eqn:est-equation} as a solution of:
\begin{equation*}
    \sum_{i=1}^n \kernel{t_i} \left[\dfrac{\kernel{t_i}(\bb{Y}(t_i) - \widehat{\beta}(\bb{x})\bb{u}_{\tau} )  }{ \Vert \kernel{t_i}(\bb{Y}(t_i) - \widehat{\beta}(x)\bb{u}_{\tau} ) \Vert } + \bb{u}_{\tau}\right] = 0,
\end{equation*}
and the null hypothesis is rejected for large values of the test statistic, i.e., if
\begin{equation*}
    \dfrac{1}{pb_n} \sup_{\bb{x} \in \chi_n} \lambda_1\left( \widehat{\bb{\Sigma}}_q(\bb{x})^{-1/2}\vert_{\Scal_p} \bb{e}_1(\bb{x})\vert_{\Scal_p}\bb{e}_1(\bb{x})\vert_{\Scal_p}\tr \widehat{\bb{\Sigma}}_q(\bb{x})^{-1/2}\vert_{\Scal_p} \right) > c_{\alpha^*}
\end{equation*}
on the relevant set $\Scal_p$ of spatial locations, where
$
    \bb{e}_1(\bb{x})\vert_{\Scal_p} = (\hat{q}_n(\tau ,\bb{x})\vert_{\Scal_p} - \widehat{\beta}(x) \bb{u}_{\tau} - \widehat{\bb{\mu}}_q(x)\vert_{\Scal_p} )
$ can be defined.

The second testing procedure with practical importance under the realm of \eqref{eqn:setup} ascribes time to a covariate through $\bb{X}(t) = t$. The hypothesis of homogeneity over time then translates into $H_0^{(2)}:\, Q(\bb{u}_{\tau}, t)/  Q( \bb{u}_{\tau}, 1) = t$, for all $t \in (0, 1]$ and $\tau \in [0, 1]$. At the significance level $\alpha^*$, $H_0^{(2)}$ is rejected if
\begin{equation*}
    \dfrac{1}{pb_n} \sup_{t \in (0, 1]} \lambda_1\left( \widehat{\bb{\Sigma}}_q(t)^{-1/2}\vert_{\Scal_p} \bb{e}_2(t)\vert_{\Scal_p}\bb{e}_2(t)\vert_{\Scal_p}\tr \widehat{\bb{\Sigma}}_q(t)^{-1/2}\vert_{\Scal_p} \right) > c_{\alpha^*}\, , 
\end{equation*}
with associated $ \bb{e}_2(t)\vert_{\Scal_p} = (\hat{q}_n(\tau, t)\vert_{\Scal_p} - t\hat{q}_n(\bb{u}_{\tau}, 1)\vert_{\Scal_p}  - \widehat{\bb{\mu}}_q(x)\vert_{\Scal_p} )$.


\section{Proofs}\label{sec:main-proofs}

This section is encompassed of the proofs of the two main theorems in this paper as well as of their ancillary results. We begin by addressing Lemma~\ref{lemma:estimating-eqn}, which consolidates the estimating equation for the proposed nonparametric quantile estimator.

\medskip
 
\begin{pfofLem}~\ref{lemma:estimating-eqn}:
    Under assumption~\ref{assum:kernelmatrix}, $\bb{K}(t) = \kernel{t}$ is a positive definite matrix for any $t \in \Tcal$. Hence, $\Phi_K(\bb{u}, \bb{Y}(t) - \bb{q}) = \Vert \bb{Y}(t) - \bb{q}\Vert_{\bb{K}(t)} + \langle \bb{u}, \bb{Y}(t) - \bb{q}\rangle_{\bb{K}(t)}$ is a strictly convex function of $\bb{q}$. Since $\widehat{\bb{q}} = \widehat{\bb{q}}_n(\bb{u}, \bb{x})$ is the minimizer of $\Mcal_{\bb{u},n}^{(p)}(\bb{q}) = \frac{1}{n}\sum_{i=1}^n \Phi_K(\bb{u}, \bb{Y}(t_i) - \bb{q})$, then for every $\bb{v} \in \R^d$,
\begin{align*}
    & \lim_{\epsilon \downarrow 0} \dfrac{1}{n\epsilon}\sum_{i=1}^n \left[ \Phi_K(\bb{u}, \bb{Y}(t_i) - \widehat{\bb{q}} + \epsilon \bb{v}) - \Phi_K(\bb{u}, \bb{Y}(t_i) - \widehat{\bb{q}} ) \right] \geqslant 0\\
    \Rightarrow \, & \dfrac{1}{n}\sum_{i=1}^n \lim_{\epsilon \downarrow 0} \dfrac{\Vert \bb{Y}(t_i) - \widehat{\bb{q}} + \epsilon \bb{v} \Vert_{\bb{K}(t_i)} - \Vert \bb{Y}(t_i) - \widehat{\bb{q}} \Vert_{\bb{K}(t_i)} }{ \epsilon } + \lim_{\epsilon \downarrow 0} \dfrac{ \langle \bb{u}, \epsilon \bb{v} \rangle_{\bb{K}(t_i)} }{\epsilon} \geqslant 0\\
    \Rightarrow \, & \dfrac{1}{n}\sum_{i=1}^{n} \left[ \dfrac{\langle \bb{Y}(t_i) - \widehat{\bb{q}}, \bb{v}\rangle_{\bb{K}(t_i)} }{ \Vert \bb{Y}(t_i) - \widehat{\bb{q}} \Vert_{\bb{K}(t_i)} } + \langle \bb{u}, \bb{v}\rangle_{\bb{K}(t_i)} \right] \geqslant 0.
\end{align*}
Using the same steps with $(-\bb{v})$ in place of $\bb{v}$ we arrive \eqref{eqn:est-equation}. Similarly, \eqref{eqn:est-equation-true} stems from \eqref{eqn:parameter}.
\end{pfofLem}


\begin{pfofThm}~\ref{thm:consistency}:
    Defining $\bbhat{q} =\widehat{\bb{q}}_n(\bb{u}, \bb{x})$ and $\bb{q}_0 = Q(\bb{u},\bb{x})$, we write the chain of inequalities:
    \begin{align*}
        \Mcal_{\bb{u}}^{(p)}(\bbhat{q}) - \Mcal_{\bb{u}}^{(p)}(\bb{q}_0)
        & \leqslant \Mcal_{\bb{u}}^{(p)}(\bbhat{q}) - \Mcal_{\bb{u}}^{(p)}(\bb{q}_0) - \Mcal_{\bb{u},n}^{(p)}(\bbhat{q}) + \Mcal_{\bb{u},n}^{(p)}(\bb{q}_0)\\
        & \leqslant 2 \sup_{\bb{q} \in \mathcal{Q}_p} \vert \Mcal_{\bb{u}, n}^{(p)}(\bb{q}) - \Mcal_{\bb{u}}^{(p)}(\bb{q}) \vert\,,
    \end{align*}
     where $\mathcal{Q}_p \subseteq \R^p$ is a set containing both $\bb{q}_0$ and the sequence of estimators $\bbhat{q}_n(\bb{u}, \bb{x})$. Hence,
    \begin{eqnarray}
    \nonumber  &     & \prob\left( \Vert \bbhat{q} - \bb{q}_0 \Vert > A p^2/n \right)\\
    \nonumber  \; &\leqslant & \prob\left( \Vert \bbhat{q} - \bb{q}_0 \Vert > A p^2/n, \,\Mcal_{\bb{u}}^{(p)}(\bbhat{q}) - \Mcal_{\bb{u}}^{(p)}(\bb{q}_0) < Bp^2/n \right)\\
    \nonumber   &   & \mbox{\hspace{4.5cm}} + \prob\left( \Mcal_{\bb{u}}^{(p)}(\bbhat{q}) - \Mcal_{\bb{u}}^{(p)}(\bb{q}_0) \geqslant Bp^2/n \right)\\
    \label{eqn:aux1}           &\leqslant & \prob\Bigl( \Vert \bbhat{q} - \bb{q}_0 \Vert > A p^2/n, \,\Mcal_{\bb{u}}^{(p)}(\bbhat{q}) - \Mcal_{\bb{u}}^{(p)}(\bb{q}_0) < Bp^2/n \Bigr)\\
     \label{eqn:aux2}  &    & \mbox{\hspace{4.5cm}} + \prob\biggl( \sup_{\bb{q} \in \mathcal{Q}_p} \vert \Mcal_{\bb{u}, n}^{(p)}(\bb{q}) - \Mcal_{\bb{u}}^{(p)}(\bb{q}) \vert   \geqslant Bp^2/2n \biggr).
    \end{eqnarray}
   

    Owing to the convexity lemma by~\citet{Pollard1991}, the objective function $\mathcal{M}_{\bb{u}}$ as part of~\eqref{eqn:parameter} is convex on $\mathcal{Q}_p$. Consequently, the sequence of conditional regression-quantile estimators $\hat{q}_n$ along $\bb{u}$ is well-defined and the true quantile $q_0$ is a well-separated minimizer of $\mathcal{M}_{\bb{u}}$, that is, there exists a $q_0 \in \mathcal{U}_p$ for which $\inf_{\bb{q}: \, \Vert \bb{q} -\bb{q}_0 \Vert \geqslant \delta }\, \left(\mathcal{M}_{\bb{u}} (\bb{q})- \mathcal{M}_{\bb{u}} (\bb{q}_0) \right) >0$, for arbitrarily small $\delta>0$. This ensures the boundedness of the first term in the above sum. In particular, since $p = o(n^{1/2})$, taking $\delta = p^2/n$ ensures that the difference $\left(\mathcal{M}_{\bb{u}} (\bb{q})- \mathcal{M}_{\bb{u}} (\bb{q}_0) \right)$ is strictly positive. It is then possible to fix $B = A\epsilon$ with arbitrarily small $\epsilon > 0$, which makes the first term equal to $0$. This takes care of \eqref{eqn:aux1}.
    %
    Consequently, the gist of the proof is in verifying the uniform convergence of  \eqref{eqn:aux2}. From the sub-Gaussian property of $\bb{Y}(t)$ together with the law of large numbers, it results that 
    \begin{equation}
        \left\vert \Mcal_{\bb{u},n}^{(p)}(\bb{q}) - \E\left[ \Vert \bb{Y}(t) - \bb{q} \Vert_{\bb{K}(t)} + \langle \bb{u}, \bb{Y}(t) - \bb{q} \rangle_{\bb{K}(t)} \right] \right\vert = \mathcal{O}_p(\sqrt{p/n}).
        \label{eqn:step-mcaln-mcalkern}
    \end{equation}
  
  In order to connect the weighted expectation affecting the kernel $\bb{K}(t)$ with $\Mcal_{\bb{u}}^{(p)}(\bb{q})$, we invoke the law of iterated expectations upon the conditioning random variable $R = \left\Vert \bb{X}(t) - \bb{x} \right\Vert$: 
    \begin{equation*}
        \E\left[ \Vert \bb{Y}(t) - \bb{q} \Vert_{\bb{K}(t)} + \langle \bb{u}, \bb{Y}(t) - \bb{q} \rangle_{\bb{K}(t)} \right]
        = \E\left[ \E\left[ \Vert \bb{Y}(t) - \bb{q} \Vert_{\bb{K}(t)} + \langle \bb{u}, \bb{Y}(t) - \bb{q} \rangle_{\bb{K}(t)} \mid R \right] \right].
    \end{equation*}

There is a one-to-one mapping between $R$ and the continuous measurable functional that is the kernel $\bb{K}(t)$. In particular, the event $R = 0$, amounts to $\bb{K}(t) = \bb{I}_p$ due to $\Kcal(0) = c_n\bb{I}_p$. Since there exists $\delta_R > 0$ such that for every $R < \delta_R$, all eigenvalues of the matrix $(\bb{K}(t) - \bb{I}_p)$ are bounded by $\sqrt{p/n}$, thus rendering valid the chain of inequalities:
    \begin{align*}
        &\left\vert \E\left[ \Vert \bb{Y}(t) - \bb{q} \Vert_{\bb{K}(t)} + \langle \bb{u}, \bb{Y}(t) - \bb{q} \rangle_{\bb{K}(t)} \right] - \E\left[ \Vert \bb{Y}(t) - \bb{q} \Vert + \langle \bb{u}, \bb{Y}(t) - \bb{q} \rangle \mid \bb{X}(t) = \bb{x} \right] \right\vert\\
        = & \left\vert \E\left[ \E\left[ \Vert \bb{Y}(t) - \bb{q} \Vert_{\bb{K}(t)} + \langle \bb{u}, \bb{Y}(t) - \bb{q} \rangle_{\bb{K}(t)} \mid R \right] \right] - \E\left[ \Vert \bb{Y}(t) - \bb{q} \Vert + \langle \bb{u}, \bb{Y}(t) - \bb{q} \rangle \mid R = 0 \right] \right\vert \\
        = & \left\vert  \E\left[ \Vert \bb{Y}(t) - \bb{q} \Vert_{\bb{K}(t)} + \langle \bb{u}, \bb{Y}(t) - \bb{q} \rangle_{\bb{K}(t)} \mid R \leqslant \delta_R \right]\prob(R \leqslant \delta_R) \right. \\
        & \qquad \left. + \E\left[ \Vert \bb{Y}(t) - \bb{q} \Vert_{\bb{K}(t)} + \langle \bb{u}, \bb{Y}(t) - \bb{q} \rangle_{\bb{K}(t)} \mid R > \delta_R \right]\prob(R > \delta_R) \right. \\
        & \qquad \left. - \E\left[ \Vert \bb{Y}(t) - \bb{q} \Vert + \langle \bb{u}, \bb{Y}(t) - \bb{q} \rangle \mid R = 0 \right] \right\vert \\
        \leqslant & \left\vert \E\left[ \Vert \bb{Y}(t) - \bb{q} \Vert_{\bb{K}(t)} - \Vert \bb{Y}(t) - \bb{q} \Vert \mid R \leqslant \delta_R \right] \right\vert \\
        & \qquad + \left\vert \E\left[ \langle \bb{u}, \bb{Y}(t) - \bb{q} \rangle_{\bb{K}(t)} - \langle \bb{u}, \bb{Y}(t) - \bb{q} \rangle \mid R \leqslant \delta_R \right] \right\vert \\
        & \qquad + \left\vert \E\left[ \Vert \bb{Y}(t) - \bb{q} \Vert_{\bb{K}(t)} + \langle \bb{u}, \bb{Y}(t) - \bb{q} \rangle_{\bb{K}(t)} \mid R > \delta_R \right] \right\vert
    \end{align*}
    
    The first term in the latter is bounded by the product of the absolute value of the maximum eigenvalue of $(\bb{K}(t) - \bb{I}_p)$ and the norm of $\Vert \bb{Y}(t) - \bb{q}\Vert$. When $R \leqslant \delta_R$, combining with the sub-Gaussian behavior of $\bb{Y}(t)$, the first term can be bounded by $\mathcal{O}(p/\sqrt{n})$. For the second term, an application of the Cauchy-Schwartz inequality combined with the fact that $\Vert \bb{u}\Vert \leqslant 1$ yields that it is also bounded by $\mathcal{O}(p/\sqrt{n})$. For the third term, note that as the bandwidth $b_n \rightarrow 0$, for sufficiently large $n$, $\delta_R > b_n$. Since the kernel $\Kcal$ has bounded support in the unit hypersphere $B^{d-1}$, it follows that $\bb{K}(t)$ can be made equal to the null matrix $\bb{0}$ in the third term. Therefore,
    \begin{equation*}
        \left\vert  \E\left[ \Vert \bb{Y}(t) - \bb{q} \Vert_{\bb{K}(t)} + \langle \bb{u}, \bb{Y}(t) - \bb{q} \rangle_{\bb{K}(t)}  \right]- \Mcal_{\bb{u}}^{(p)}(\bb{q}) \right\vert = \mathcal{O}(p/\sqrt{n})
    \end{equation*}
    
    Combining this with~\eqref{eqn:step-mcaln-mcalkern}, we obtain that for each fixed $\bb{q} \in \R^p$, 
    \begin{equation*}
        \Vert \Mcal_{\bb{u}, n}^{(p)}(\bb{q}) - \Mcal_{\bb{u}}^{(p)}(\bb{q}) \Vert = \mathcal{O}_p(\max\{ p^2/n, p/\sqrt{n} \}) = \mathcal{O}_p(p/\sqrt{n}),
    \end{equation*} 
    \noindent where the last equality follows from the initial assumption that $p = o(n^{1/2})$. The uniform convergence holds if we can show that $\mathcal{Q}_p$ is a compact set. Thus, it is enough to show that the sequence of estimators $\bb{q}_n$ minimizing the sequence of objective functions $\Mcal_{\bb{u}, n}^{(p)}(\bb{q})$ are uniformly bounded. Without the loss of generality, consider that $0$ belongs to the convex hull of the datapoints $\{ \bb{Y}(t_i) : i = 1, 2, \dots n\}$, otherwise we can translate all datapoints suitably. By applying the sub-Gaussian property of $\bb{Y}(t_i)$ and union bound, it follows that there exists some $c > 0$ such that
    \begin{equation*}
        \prob\left( \sup_{1 \leqslant i \leqslant n} \Vert \bb{Y}(t_i)\Vert > M \right) \leqslant ne^{-cp}.
    \end{equation*}
    
    As $p^{-1}\log(n) = \mathcal{O}(1)$, by choosing sufficiently large $p \geqslant C \log(n)$ for some absolute constant $C$, 
    we can now make $\Mcal_{\bb{u}, n}^{(p)}(\bb{0}) < M$ with probability arbitrarily close to $1$, by bounding each $\bb{Y}(t_i)$. Also, since $\Vert \bb{u}\Vert < 1$, there exists a $\delta \in (0, 1)$ such that $\Vert \bb{u}\Vert < (1-\delta)$, and a vector $\bb{v}$ such that $\Vert \bb{v}\Vert > 2M\delta^{-1}$. Consider the objective function at a point $\bb{Y}(t_j) + \bb{v}$ for any $j = 1, 2, \dots n$;
    \begin{align*}
        \Mcal_{\bb{u}, n}^{(p)}(\bb{Y}(t_j) + \bb{v}) 
        & = \dfrac{1}{n} \sum_{i=1}^n \left[ \Vert \bb{Y}(t_i) - \bb{Y}(t_j) - \bb{v} \Vert_{\Kcal(t_i)} + \langle \bb{u}, \bb{Y}(t_i) - \bb{Y}(t_j) - \bb{v}  \rangle_{\Kcal(t_i)} \right]\\
        & \geqslant \Vert \bb{v}\Vert/2 - (1 - \delta)\Vert \bb{v}\Vert/2, \mbox{ (by Cauchy-Schwartz inequality)}\\
        & \geqslant M \geqslant \Mcal_{\bb{u}, n}^{(p)}(\bb{0})
    \end{align*}
    
    By general positioning of these vectors, $\{ \bb{Y}(t_j) + \bb{v}, \bb{Y}(t_j) - \bb{v} : j = 1, \dots n\}$ do not lie in a subspace with a dimension strictly less than $p$. Combining this with the convex nature of the objective function $\Mcal_{\bb{u}, n}^{(p)}(\bb{q})$, it follows that along the boundary of the convex hull of these points also, the value of the objective function is more than $M$. Using continuity of the objective function, we find that there is a $(p-1)$-dimensional hypersphere, independent of the choice of $n$, containing the origin $\bb{0}$ such that the $\Mcal_{\bb{u}, n}^{(p)}(\bb{q}) > \Mcal_{\bb{u}, n}^{(p)}(\bb{0})$ for any $\bb{q}$ lying on the boundary of that hypersphere for any $n$. Now it follows from the strict convexity of each $\Mcal_{\bb{u}, n}^{(p)}(\bb{q})$ that they have a local minimum within that hypersphere. Choosing the $\bbhat{q}_n$ as those local minima establishes the boundedness of them.     
\end{pfofThm}


\begin{pfofThm}~\ref{thm:normal-dist}:
 We note that under the conditions of the theorem, Propositions~\ref{prop:first-div-dist}-\ref{prop:third-div-bound} are valid. We start by employing Taylor's expansion of $\nabla \Mcal_{\bb{u}, n}^{(p)}(\bb{q})$ about the true parameter $\bb{q}_0 := Q(\bb{u}, \bb{x})$ while evaluated at $\widehat{\bb{q}} := \widehat{\bb{q}}_n(\bb{u},\bb{x})$, to obtain:
    \begin{align*}
        \bb{0} & = \nabla \Mcal_{\bb{u}, n}^{(p)}(\widehat{\bb{q}})\\
        & = \nabla \Mcal_{\bb{u}, n}^{(p)}(\bb{q}_0) + \nabla^2 \Mcal_{\bb{u}, n}^{(p)}(\bb{q}_0)(\widehat{\bb{q}} - \bb{q}_0) + \mathcal{O}\Bigl(\Vert \widehat{\bb{q}} - \bb{q}_0\Vert^2 \left\Vert \nabla^3\Mcal_{\bb{u}, n}^{(p)}(\bb{q}) \right\Vert \Bigr)
    \end{align*}
    
    Denoting $\nabla^{i} \Mcal_{\bb{u}, n}^{(p)}(\bb{q}_0) = \Delta_{i}$, we have, with $\delta_n = p/\sqrt{n}$, that
    \begin{equation*}
        a_n\left( \widehat{\bb{q}} - \bb{q}_0 \right) = -a_n \Delta_{2}^{-1} \Delta_1 + \mathcal{O}(\Vert \Delta_2^{-1}\Vert \delta_n^2 / pb_n).
    \end{equation*}
   
    Here, we use the consistency result of $\widehat{\bb{q}}$ coupled with proposition~\ref{prop:third-div-bound} to bound $\Vert \Delta_3\Vert$. By proposition~\ref{prop:second-div-mat}, it follows that for sufficiently large $n$,
    \begin{equation*}
        \left\Vert \Delta_2^{-1} - \bb{\Psi}(\bb{q}_0)^{-1}\right\Vert \leqslant \dfrac{2}{\Vert \bb{\Psi}(\bb{q}_0) \Vert^2} \left\Vert \bb{\Psi}(\bb{q}_0) - \Delta_{2}\right\Vert \leqslant \dfrac{Cp^2}{n\Vert \bb{\Psi}(\bb{q}_0) \Vert^2}.
    \end{equation*}
    
    Therefore,
    \begin{align*}
        & \quad \left\vert \E\left( h\left( a_n\left( \widehat{\bb{q}} - \bb{q}_0 \right) \right) \right) - \E\left( h\left( \bb{\Psi}^{-1}(\bb{q}_0) Z(\bb{q}_0) \right)  \right) \right\vert \\
        \leqslant & \quad \left\vert \E\left( h\left( -a_n \Delta_{2}^{-1} \Delta_1 + \mathcal{O}(\Vert \Delta_2^{-1}\Vert p/nb_n ) \right) \right) - \E\left( h\left( \bb{\Psi}^{-1}(\bb{q}_0) Z(\bb{q}_0) \right)  \right) \right\vert \\
        \leqslant & \quad \left\vert \E\left( h\left( -a_n \Delta_{2}^{-1} \Delta_1 + \mathcal{O}(\Vert \Delta_2^{-1}\Vert p/nb_n ) \right) \right) - \E\left( h\left( -a_n\bb{\Psi}^{-1}(\bb{q}_0) \Delta_1 \right)  \right) \right\vert + \\
        & \qquad \qquad \left\vert \E\left( h\left( -a_n\bb{\Psi}^{-1}(\bb{q}_0) \Delta_1 \right) \right) - \E\left( h\left( \bb{\Psi}^{-1}(\bb{q}_0) Z(\bb{q}_0) \right)  \right) \right\vert\\
        \leqslant & \quad \E\left[ 
        \left\Vert h\dash(\widetilde{\bb{q}}) \right\Vert \left( \mathcal{O} \left(\Vert \Delta_2^{-1}\Vert  \frac{p}{nb_n} \right) + \vert a_n\vert \Vert \Delta_1 \Vert \left\Vert \Delta_2^{-1} - \bb{\Psi}(\bb{q}_0)^{-1} \right\Vert  \right) \right] + \mathcal{O}(p^2/n)\, ,
    \end{align*}
    where $\widetilde{\bb{q}}$ is some vector lying between $-a_n\bb{\Psi}^{-1}(\bb{q}_0) \Delta_1$ and  $-a_n \Delta_{2}^{-1} \Delta_1$, ensured by the mean value theorem. Here, we use the fact that the function $g(x) := \bb{\Psi}^{-1}(\bb{q}_0) f(x)$ is also an element of $\mathcal{H}$, hence $\left\vert \E\left( g\left( -a_n \Delta_1 \right) \right) - \E\left( g\left( Z(\bb{q}_0) \right)  \right) \right\vert$ is $\mathcal{O}(p^2/n)$ by virtue of~\ref{prop:first-div-dist}. On the other hand, due to the choice of $\Hcal$ and $h \in \Hcal$, the quantity $\left\Vert h\dash(\widetilde{\bb{q}}) \right\Vert$ remains bounded. 
    Noting that $\left\Vert \bb{\Psi}(\bb{q}_0) \right\Vert$ is positive definite and has bounded entries,  its eigenvalues are of order $\mathcal{O}(\sqrt{p})$. This implies that the first term is at most $\mathcal{O}(p^{17/8}/n)$, whence
    \begin{equation*}
        \left\vert \E\left( h\left( a_n\left( \widehat{\bb{q}} - \bb{q}_0 \right) \right) \right) - \E\left( h\left( \bb{\Psi}^{-1}(\bb{q}_0) Z(\bb{q}_0) \right)  \right) \right\vert
        = \mathcal{O}(p^{17/8}/n).
    \end{equation*} 
    
    Since this bound does not depend on any given $h \in \Hcal$, the above conclusion holds with the supremum over $h \in \Hcal$. Finally, the choice of $\bb{u}$ is restricted within the compact set $\mathcal{B}_p$, giving way to a further supremum on this set. 
\end{pfofThm}


\section{Forecasting energy demand}
\label{Sec:real-data}

In order to experimentally verify the good overall performance of the proposed quantile estimator, we conduct a simulation study that encompasses two distinct setups: one without any explicit spatial information and another setup with proper spatial information included. In both cases, we find that the proposed estimator achieves reasonable prediction accuracy. The simulation results are compiled in Appendix B featuring in the supplementary material.

In this section, the proposed quantile regression methodology is applied to produce inferences on the Thames Valley Vision (TVV) data, collected by Scottish and Southern Electricity Network (SSEN), the primary industrial partner of the TVV project funded by UK gas and electricity regulator Ofgem through the Low Carbon Networks Fund and Network Innovation Competition. The dataset handled here has been partly studied in~\cite{jacob2020forecasting}. The project's overall aim was to evaluate a typical low voltage network using monitoring in households and substations in order to simulate future demand scenarios.

\begin{figure}[t]
    \centering
    \includegraphics[scale=0.42]{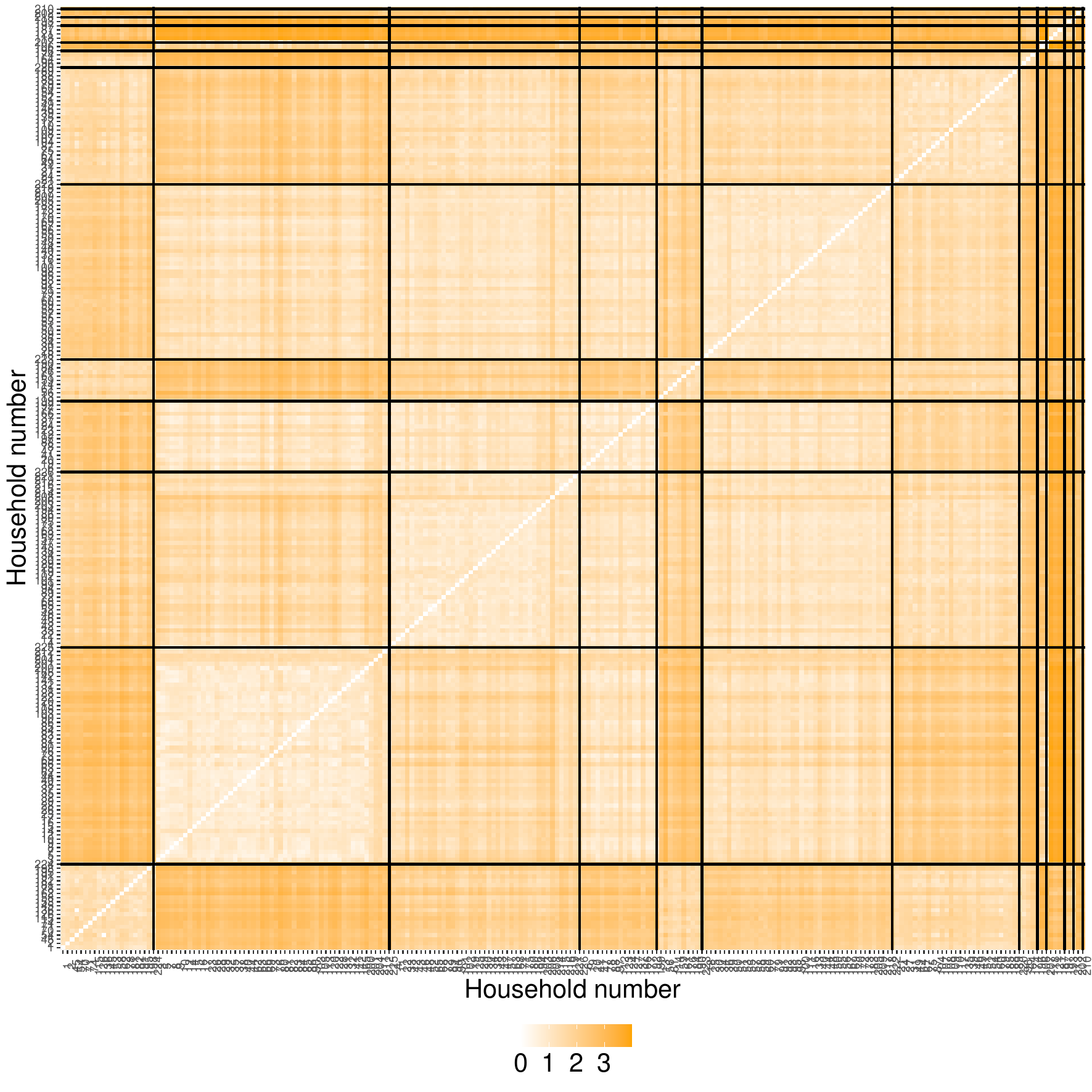}
    \caption{Heatmap of DTW distances (transformed to log-scale) between average daily household profiles, arranged according to their assigned clusters.}
    \label{fig:heatplot-log}
\end{figure}

The dataset under study consists of the energy profiles for $226$ households with half-hourly resolution, between 19th July 2015 and 20th September 2015, thus spanning $9$ weeks of data. In our analysis, the first $8$ weeks of data are used as a training set whereas our proposed estimator is employed to predict different quantiles for last week's electricity demand. Given the temporal resolution of the data, the number of observations for every location in the training set is $n = 2688$. 
In line with the theory underpinning the quantile estimation introduced in the present paper (see Theorem~\ref{thm:consistency}),  the number of spatial points $p=226$ far exceeds the benchmark of $\sqrt{n} \approx 52$. This provides opportunity to split the households into clusters of size not larger than $52$ based on their average daily demand (ADD) profiles. To determine these clusters, we calculated the pairwise dynamic time warping (DTW) distance between every two household's ADD profiles using the FastDTW algorithm from~\cite{gold2018dynamic} with Hierarchical clustering on the households. \Cref{fig:heatplot-log} displays the heatmap for the pairwise DTW distances (log-transformed) between households. Through this procedure, the households were assigned to clusters $j= 1, 2, \ldots, 12 $ of sizes 21, 52, 42, 17, 10, 42, 28, 4, 2, 4, 2 and 2, respectively. 

We find that the pairwise within-cluster distances tend to be smaller than the between-cluster distances, which is reflected in the lighter shaded blocks containing the reverse diagonal of the heatmap. Hence, it is reasonable to assume that electricity demand profiles of two households residing in two different clusters are independent or weakly dependent (the latter is not a deterrent to our methods). Consequently, we applied our proposed quantile estimation to each cluster separately. In a nutshell, our statistical analysis revolves broadly around two goals:

\begin{itemize}
    
    \item[(i)] the estimation of relevant quantiles for each cluster of households and prediction of the quantiles for profiles in the $9$-th week across all $226$ households. As covariates, we take day of the week and hour of the day, as well as a trend over time (not necessarily monotonic).
    
    \item[(ii)] the statistical evaluation of homogeneity of the effects of the day and hour covariates for different households. In addition, discerning significance of a potential trend over time in the quantiles by relying on  the methodology described in \Cref{sec:ht}.
    
\end{itemize}  

Throughout this analysis, for ease of implementation and interpretation, we fix the levels of 0.5, 0.75, 0.8, 0.9, 0.95, 0.99 across all the cases examined.

\subsection{One-week-ahead quantile estimation}

\begin{figure}[!th]
    \centering
    \includegraphics[scale=0.45]{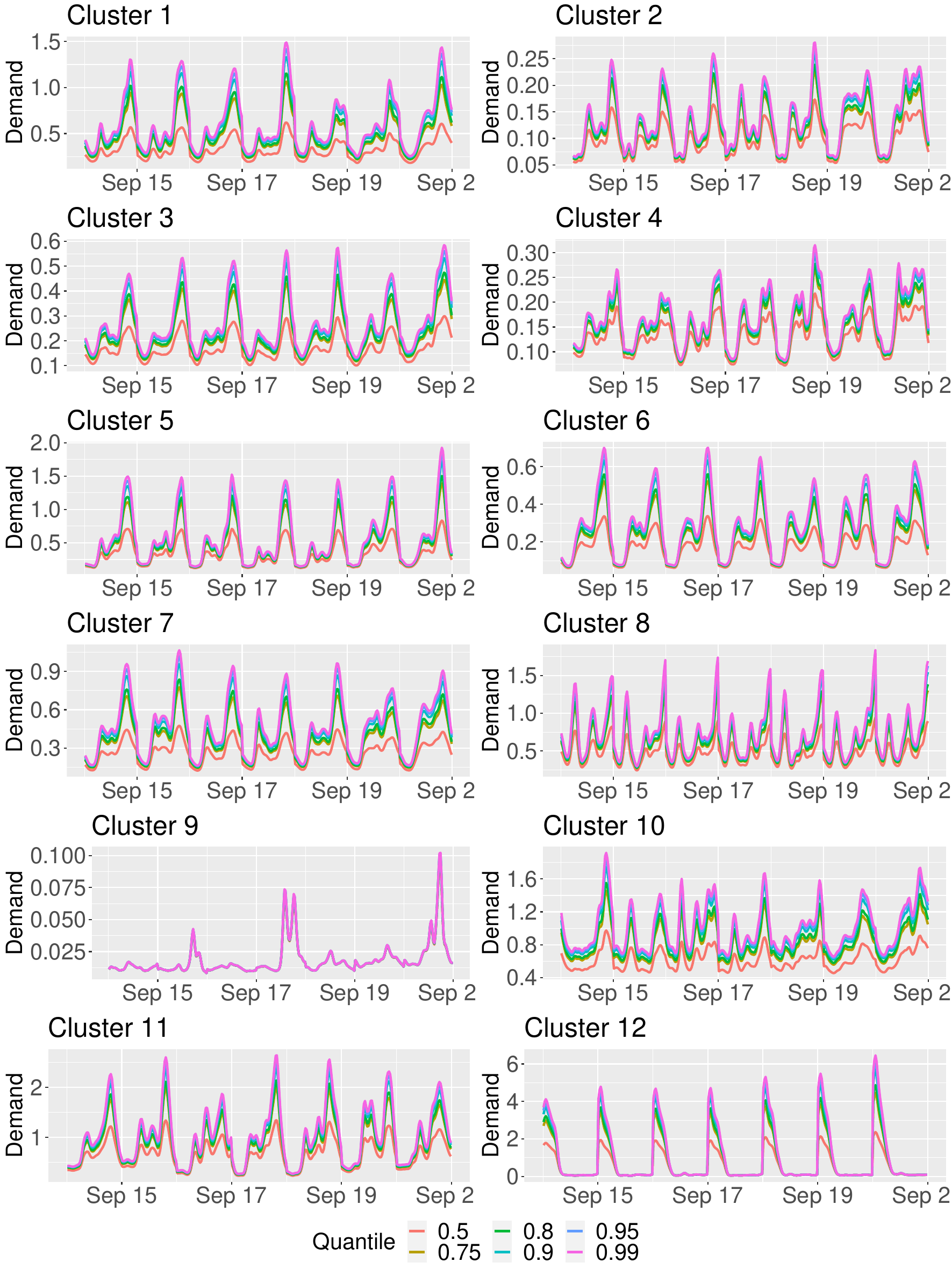}
    \caption{Predicted quantile curves for electricity demand during the 9th week out of 9 weeks of daily profile data, for cluster $j= 1, 2, \ldots, 12$ of size 21, 52, 42, 17, 10, 42, 28, 4, 2, 4, 2 and 2.}
    \label{fig:estimate-cluster}
\end{figure}

For tackling the estimation problem described in (i), we estimated quantile functions from the training data relating the electricity demand profiles per cluster. The one-week ahead predicted quantiles are displayed in~\Cref{fig:estimate-cluster}. The plots seem to indicate that the estimated quantiles enjoy a smoothness property due to the continuous and smooth nature of the kernel function adopted in the estimator. There is no apparent quantile crossing issue as well. Unsurprisingly, two quantile curves draw closer when the general trend in demand is approximately constant, but on certain time of the day quantile curves associated with the larger probabilities shoot away from the median level, in what can be identified as peak hours. The systematic cyclical pattern inherent to electricity demand due to the daily day-night cycle is also captured in the prediction. In particular, clusters 2, 4, 8 and 12 indicate high electricity loads during the weekends. While for most clusters, the quantile curves provide a satisfactory description of how a proper probability distribution would spread, the quantile curves for cluster 9 crop up very tightly together. This is mainly because there are only two households (with numbers 106 207) in this cluster, making the number of spatial points clearly insufficient for inference on the true range of the quantiles and not the least of their temporal evolution. On the same note, the pattern of the cyclical nature of the electricity demand for Cluster 12 is in stark contrast with other clusters, where peak values tend to occur at the beginning of the day. The households in this cluster are  202 and 210.

\subsection{Hypotheses testing}\label{Sec:realdata-ht}

Before getting underway, it is important to caution the reader  that the two testing procedures illustrated in this section are computationally challenging. Since computational efficiency is not the main focus of this paper, we flag this up as a pertinent aspect for further improvement and limit the present study to two clusters. The selected clusters 2 and 12 to this undertaking are the two extreme cases in terms of size, the second cluster enclosing the maximum allowed of 52 households while the 12th cluster is made up of only two. \Cref{tab:hypothesis-testing} contains the $p$-values of the two tests of homogeneity introduced in Section \ref{sec:ht} conducted for several quantile levels, $\tau= 0.5, 0.75, 0.8, 0.9, 0.95, 0.99$.

\begin{table}[t]
    \centering
    \caption{Computed $p$-values for the results of the two tests for assessing homogeneity.}
    \label{tab:hypothesis-testing}
    \begin{tabular}{cccc}
    \toprule
    Test & Quantile & Cluster 2 & Cluster 12 \\
    \midrule
    Homogeneity of & 0.5 & $8.07\times 10^{-6}$ & $1.05 \times 10^{-6}$ \\
    effect of covariates & 0.75 & $6.45\times 10^{-6}$ & $1.04\times 10^{-6}$ \\
    & 0.9 & $8.10 \times 10^{-6}$ & $1.05 \times 10^{-6}$ \\
    & 0.95 & $7.96\times 10^{-6}$ & $1.05 \times 10^{-6}$ \\
    & 0.99 & $8.07\times 10^{-6}$ & $1.05 \times 10^{-6}$ \\
    \midrule
    Homogeneity of & 0.5 & $2.79\times 10^{-7}$ & $1.88 \times 10^{-8}$ \\
    effect of time & 0.75 & $2.03\times 10^{-7}$ & $1.88 \times 10^{-8}$ \\
    & 0.9 & $2.78\times 10^{-7}$ & $1.88 \times 10^{-8}$ \\
    & 0.95 & $2.79\times 10^{-7}$ & $1.88 \times 10^{-8}$ \\
    & 0.99 & $2.79\times 10^{-7}$ & $1.88 \times 10^{-8}$ \\
    \bottomrule
    \end{tabular}
\end{table}

It is clear that all $p$-values yielded are highly significant, leading to the rejection of the null hypothesis of homogeneity in all cases. Interestingly, this finding is unchallenged by the size the clusters and is also consistent across quantile levels. Therefore, the is sufficient evidence to conclude that the covariates have significantly different effects for different household's demands. There is also significant evidence of a temporal trend in the quantiles of all levels considered. Given the magnitude of the $p$-values, these conclusions hold for any nominal level of test, i.e., both null hypotheses are rejected notably at the minimum of the usual significance levels ($\alpha= 1\%$).

\section*{Acknowledgements} {

Cl\'audia Neves gratefully acknowledges support from EPSRC-UKRI Innovation Fellowship grant EP/S001263/2. Her work is also partly supported by CEAUL, Faculty of Science, University of Lisbon, DOI: 10.54499/UIDB/00006/2020,
\url{https://doi.org/10.54499/UIDB/00006/2020}.}

\section{Supplementary documentation}

The supplementary material is comprised of two sections, namely: Appendix A with the proof for Proposition~\ref{prop:P1P3} in full;  Appendix B containing simulation results based on a comprehensive set of numerical experiments.

%
\bibliographystyle{apalike} 
\bibliography{QRreferences.bib}      


\end{document}


\begin{frontmatter}
\title{Supplementary material for ``Nonparametric quantile regression for spatio-temporal processes''}
\runtitle{Spatio-temporal quantile regression}

\begin{aug}
\author[A]{\fnms{Soudeep}~\snm{Deb}\ead[label=e2]{soudeep@iimb.ac.in}\orcid{0000-0003-0567-7339}}
\author[B]{\fnms{Cl\'audia}~\snm{Neves}\ead[label=e3]{claudia.neves@kcl.ac.uk}\orcid{0000-0003-1201-5720}}
\and
\author[C]{\fnms{Subhrajyoty}~\snm{Roy}\ead[label=e1]{roysubhra98@gmail.com}\orcid{0000-0003-0909-5635}}

\runauthor{Deb, Neves, Roy}


\address[A]{Decision Sciences,
Indian Institute of Management Bangalore\printead[presep={,\ }]{e2}}

\address[B]{Department of Mathematics,
King's College London\printead[presep={,\ }]{e3}}

\address[C]{Applied Statistics Division, Indian Statistical Institute Kolkata\printead[presep={,\ }]{e1}}

\end{aug}

\begin{abstract}
This supplementary material document consists of two sections: Appendix A with the proof for Proposition~\ref{F1-prop:P1P3} in full;  Appendix B containing simulation results based on a comprehensive set of numerical experiments.
\end{abstract}



\end{frontmatter}

\setcounter{equation}{27} 

\appendix
\section{Proof of Proposition~\ref{F1-prop:P1P3}}\label{appA}

\subsection{Proof of~\ref{F1-prop:first-div-dist}}\label{proof:first-div-dist}
We write:
\begin{multline}\label{Aux}
 -a_n \nabla \Mcal_{\bb{u},n}^{(p)}(\bb{q}_0)
    = \dfrac{a_n}{n} \sum_{i=1}^n \bb{K}(t_i) \left(\dfrac{\bb{\eta}(t_i)}{\Vert \bb{\eta}(t_i)\Vert} + \bb{u} \right) + \dfrac{a_n}{n} \sum_{i=1}^n \bb{K}(t_i) \bb{V}_2(t_i)\bb{V}_1^{1/2}(t_i) \bb{\xi}(t_i)\\
    + \dfrac{a_n}{n}\sum_{i=1}^n \mathcal{O}_p(\Vert \bb{K}(t_i) \Vert \Vert \bb{\eta}(t_i)\Vert^{-3} \Vert \bb{V}_1^{1/2}(t_i) \Vert \Vert \bb{\xi}(t_i) \Vert^2).
\end{multline}

Let us begin by dealing with the third term. Due to the sub-Gaussian nature of $\bb{\xi}(t)$ for any $t \in \Tcal$, $\Vert \bb{\xi}(t) \Vert = \mathcal{O}_p(\sqrt{p})$, $\Vert \bb{\xi}(t) \Vert^2 = \mathcal{O}_p(p + \sqrt{p})$ and $\Vert \bb{\xi}(t)\Vert^3 = \mathcal{O}_p(p\sqrt{p} + p)$, with probability greater than or equal to $(1 - e^{-cp})$. Since each element of the kernel matrix is an $\mathcal{O}(b_n)$, its norm is bounded by $\mathcal{O}(\sqrt{p}b_n)$ almost surely~\citep{tao2012topics}. Thus,
\begin{align*}
    \dfrac{a_n}{n}\sum_{i=1}^n \mathcal{O}_p(\Vert \bb{K}(t_i) \Vert \Vert \bb{\eta}(t_i)\Vert^{-3} \Vert \bb{V}_1^{1/2}(t_i)\Vert \Vert \bb{\xi}(t_i) \Vert^2)
    &= \mathcal{O}\left( \dfrac{a_n b_n^2 p^{3/4} (p+\sqrt{p}) }{b_n^3 (p\sqrt{p} - p)^2 } \right)\\
    &= \mathcal{O}\left( b_n^{-2}p^{-9/4} \right).
\end{align*}
which vanishes to $0$ almost surely because we assume as part of~\ref{F1-assum:alpha-beta} that $p^{9/8}b_n \rightarrow \infty$ as $n \rightarrow \infty$ (see also text underneath Theorem~\ref{F1-thm:normal-dist}).


In relation to the first term, we note that $\left\Vert {\bb{\eta}(t_i)}/{\Vert \bb{\eta}(t_i)\Vert} \right\Vert \leqslant 1$ and $\Vert \bb{u} \Vert < 1$. Consequently, we can rewrite it as
\begin{align*}
    & \quad \dfrac{a_n}{n}\sum_{i=1}^n \bb{K}(t_i) \left(\dfrac{\bb{\eta}(t_i)}{\Vert \bb{\eta}(t_i)\Vert} + \bb{u} \right)\\
    = & \quad \dfrac{a_n}{n}\sum_{i=1}^n \bb{K}(t_i) \left( \dfrac{\bb{\eta}(t_i)}{\Vert \bb{\eta}(t_i)\Vert} - \E_{\bb{Y}(t) \mid \bb{X}(t) = \bb{x}}\left( \dfrac{\bb{Y}(t) - \bb{q}_0}{\Vert \bb{Y}(t) - \bb{q}_0 \Vert} \right) \right) + \\
    & \quad \quad \quad \dfrac{a_n}{n}\sum_{i=1}^n \bb{K}(t_i) \E_{\bb{Y}(t) \mid \bb{X}(t) = \bb{x}}\left( \dfrac{\bb{Y}(t) - \bb{q}_0}{\Vert \bb{Y}(t) - \bb{q}_0 \Vert} + \bb{u} \right)\\
    = & \quad \dfrac{a_n}{n}\sum_{i=1}^n \bb{K}(t_i) \left( \dfrac{\bb{\eta}(t_i)}{\Vert \bb{\eta}(t_i)\Vert} - \E_{\bb{Y}(t) \mid \bb{X}(t) = \bb{x}}\left( \dfrac{\bb{Y}(t) - \bb{q}_0}{\Vert \bb{Y}(t) - \bb{q}_0 \Vert} \right) \right), \\
    & \quad \quad \quad \mbox{(the second summand becomes $0$ via Lemma~\ref{F1-lemma:estimating-eqn})}\\
    = & \quad \dfrac{a_n}{n}\sum_{i=1}^n \bb{K}(t_i) \left( \dfrac{\bb{\eta}(t_i)}{\Vert \bb{\eta}(t_i)\Vert} - \dfrac{\bb{\mu}(\bb{X}(t)) - \bb{q}_0}{\Vert \bb{\mu}(\bb{X}(t)) - \bb{q}_0 \Vert} + \mathcal{O}\left( p^{-1/2} \right) \right)\\
    = & \quad \dfrac{a_n}{n}\sum_{i=1}^n \bb{K}(t_i) \left( \dfrac{\bb{\eta}(t_i)}{\Vert \bb{\eta}(t_i)\Vert} - \dfrac{\bb{\mu}(\bb{X}(t)) - \bb{q}_0}{\Vert \bb{\mu}(\bb{X}(t)) - \bb{q}_0 \Vert} \right) + \mathcal{O}(b_n^{-1}p^{-3/2}),
\end{align*}
where we again apply Taylor series decomposition for $\E_{\bb{Y}(t) \mid \bb{X}(t) = \bb{x}}\left( \dfrac{\bb{Y}(t) - \bb{q}_0}{\Vert \bb{Y}(t) - \bb{q}_0 \Vert} \right)$ and bound the norms using standard sub-Gaussian tail bounds. Clearly, whenever $b_n^{-2}p^{-9/4} \rightarrow 0$, it readily follows that $b_n^{-1}p^{-3/2} \rightarrow 0$ as $n \rightarrow \infty$. In view of \eqref{F1-eqn:keydef-eta}, the first term has limit $\bb{\eta}_t(\bb{q}_0)$.  Furthermore, the variance
 of $\sum_{i=1}^n \bb{K}(t_i) (\bb{\eta}(t_i)/\Vert \bb{\eta}(t_i)\Vert + \bb{u})$ is bounded:
\begin{multline*}
    \left\Vert \var\left( \dfrac{a_n}{n} \sum_{i=1}^n \bb{K}(t_i) \left( \dfrac{\bb{\eta}(t_i)}{\Vert \bb{\eta}(t_i) \Vert} + \bb{u}\right) \right) \right\Vert
    \leqslant \dfrac{a_n^2}{n^2} \sum_{i=1}^n \Vert \bb{K}(t_i)\Vert^2 \left\Vert \var\left( \dfrac{\bb{\eta}(t_i)}{\Vert \bb{\eta}(t_i) \Vert} \right) \right\Vert\\
    = \mathcal{O}\left( \dfrac{a_n^2}{n} (p+\sqrt{p})b_n^2 \right)
    = \mathcal{O}\left( (np)^{-1} \right),
\end{multline*}
where $\left\Vert \var\left( \bb{\eta}(t_i) / \Vert \bb{\eta}(t_i)\Vert \right) \right\Vert = \mathcal{O}(1)$ since the norm of the random variable involved is uniformly bounded by $1$. Clearly, as $n \rightarrow \infty$, the variance tends to $0$ which, given the assumptions in the proposition, implies
\begin{equation*}
    \left\Vert \frac{a_n}{n} \sum_{i=1}^n \bb{K}(t_i) \left(\dfrac{\bb{\eta}(t_i)}{\Vert \bb{\eta}(t_i)\Vert} + \bb{u} \right) - \bb{\eta}_t(\bb{q}_0) \right\Vert \xrightarrow{P} 0.
\end{equation*}

Next, we consider the second term. We note that, $\bb{\eta}(t_i)$, $\bb{V}_1(t_i)$ and $\bb{V}_2(t_i)$ are all predictable with respect to the filtration $\{ \mathcal{G}_{t} : t \in \Tcal \}$. Hence, $\bb{K}(t_i)\bb{V}_2(t_i)\bb{V}_1^{1/2}(t_i)\bb{\xi}(t_i)$ is a martingale difference with respect to the same filtration. Letting $\bb{\zeta}(t_i) := ({a_n}/{n}) \bb{K}(t_i)\bb{V}_2(t_i)\bb{V}_1^{1/2}(t_i)\bb{\xi}(t_i)$, we have that
\begin{equation*}
    \bb{\Sigma}_{\zeta, i}^2
    = \E\left[ \bb{\zeta}(t_i)\bb{\zeta}(t_i)\tr \mid \Gcal_{t_i-} \right]
    = \frac{a_n^2}{n^2} \bb{K}(t_i) \bb{V}_2(t_i)\bb{V}_1(t_i)\bb{V}_2(t_i)\tr \bb{K}(t_i)
\end{equation*}
and, in view of \eqref{F1-eqn:keydef-omega}, 
\begin{equation*}
    \Delta_n := \left\Vert \sum_{i=1}^n \bb{\Sigma}_{\zeta,i}^2 - \bb{\Omega}_t(\bb{q}_0) \right\Vert =o(1).
\end{equation*}

Let $h \in \mathcal{H}$, the set of three-times differentiable functions with bounded first, second and third derivative, and let $c_2$ and $c_3$ denote the upper bound of the second and third derivative of $f$ respectively.  Then, application of the quantitative central limit theorem for the multidimensional martingale~\citep{belloni2018high} ensures that
\begin{equation*}
    \left\Vert \E\left[ h\left(\dfrac{a_n}{n} \sum_{i=1}^n \bb{\zeta}(t_i) \right) \right] - \E \left[ h(\bb{Z}^{\ast}(\bb{q}_0)) \right] \right\Vert  \leqslant 2c_2 \Delta_n + c_3 \sum_{i=1}^n \E\left[ \left\Vert \bb{\zeta}(t_i) \right\Vert^3 \right],
\end{equation*}
where $\bb{Z}^\ast(\bb{q}_0)$ is a mean-zero Gaussian vector with dispersion matrix $\bb{\Omega}_t(\bb{q}_0)$. Noting that $\Delta_n \rightarrow 0$, we may now write
\begin{equation*}
    \sum_{i=1}^n \E\left[ \left\Vert \bb{\zeta}(t_i) \right\Vert^3 \right]
    = \mathcal{O}\left(\dfrac{a_n^3}{n^3}\sum_{i=1}^n (p\sqrt{p}+p)^3 b_n^3 \right)
    = \mathcal{O}(n^{-2}p^{3/2}).
\end{equation*}

Since the primary operative assumption is that $p=p_n$ is an intermediate sequence of positive integers such that $p \rightarrow \infty$ and $p/n = o(1)$, then $p^{3/2}/n^2 =o(1)$.


Combining all the terms dealt in the above through the use of Slutsky's theorem, we have for any fixed $\bb{u} \in B^{p-1}$ and for sufficiently large $n$, that
\begin{align*}
   & \sup_{h \in \Hcal} \left\Vert \E\left[f \left(a_n (\widehat{q}_n(\bb{u},\bb{x}) - Q(\bb{u},\bb{x}) \right)\right] - \E\left[ h\left( -\bb{\Omega}_t^{-1/2}(\bb{q}_0)\bb{Z}(\bb{q}_0) \right) \right] \right\Vert\\
   = & \; \mathcal{O}\left(\max\bigl\{p^2/n, (b_n p^{9/8})^{-1} \bigr\} \right),
\end{align*}
where  $\bb{Z}(\bb{q}_0)$ is provided in the statement of proposition~\ref{F1-prop:first-div-dist} (cf.~\ref{F1-assum:alpha-beta}). Finally, we apply a standard $\epsilon$-net argument \citep[see Section 2.3.1 of][]{tao2012topics} with union bounds over the compact subset $\mathcal{B}$ of the unit ball $B^{p-1}$ to take a supremum over the choice of $\bb{u}$, which transforms the above into the uniform convergence ascertained in Proposition~\ref{F1-prop:first-div-dist}.

\subsection{Proof of~\ref{F1-prop:second-div-mat}}\label{proof:second-div-mat}

First, we express the term $\nabla^2 \Mcal_{\bb{u},n}^{(p)}(\bb{q})$ by computing the second order derivative of the objective function $\Mcal_{\bb{u},n}^{(p)}(\bb{q})$ as follows
\begin{equation*}
    \nabla^2 \Mcal_{\bb{u},n}^{(p)}(\bb{q}) 
    = \dfrac{1}{n}\sum_{i=1}^n \bb{K}(t_i) \left[ \dfrac{\bb{I}_p}{\left\Vert  \bb{Y}(t_i) - \bb{q} \right\Vert_{\bb{K}(t_i)} } 
 - \dfrac{\bb{K}(t_i)(\bb{Y}(t_i) - \bb{q})(\bb{Y}(t_i) - \bb{q})\tr \bb{K}(t_i)}{\left\Vert  \bb{Y}(t_i) - \bb{q} \right\Vert^3_{\bb{K}(t_i)}} \right] \bb{K}(t_i).
\end{equation*}

We perform an M-R decomposition technique on $\nabla^2 \Mcal_{\bb{u},n}^{(p)}(\bb{q})$ as in~\cite{zhao2008confidence} and rewrite 
\begin{equation*}
    \nabla^2 \Mcal_{\bb{u},n}^{(p)}(\bb{q}) = M_{n,p} + R_{n,p}
\end{equation*}
where 
\begin{align*}
    \bb{a}(t_i) & = \bb{K}(t_i) (\bb{Y}(t_i) - \bb{q}),\\
    R_{n,p} & = \dfrac{1}{n}\sum_{i=1}^n \bb{K}(t_i) \E\left[ \dfrac{\bb{I}_p}{\Vert \bb{a}(t_i)\Vert } - \dfrac{\bb{a}(t_i) \bb{a}(t_i)\tr}{\Vert \bb{a}(t_i)\Vert^3 }  \mid \Gcal_{t_i-1} \right] \bb{K}(t_i),\\
    M_{n,p} & = \dfrac{1}{n}\sum_{i=1}^n \bb{K}(t_i) \left[ \dfrac{\bb{I}_p}{\Vert \bb{a}(t_i)\Vert } - \dfrac{\bb{a}(t_i) \bb{a}(t_i)\tr}{\Vert \bb{a}(t_i)\Vert^3 } \right] \bb{K}(t_i) - R_{n,p}.
\end{align*}

We note that $M_{n,p}$ is the partial average of a martingale difference sequence with respect to the filtration $\{ \Gcal_{t} : t \in \Tcal \}$. Applying the quantitative central limit theorem for multidimensional martingale~\citep{belloni2018high}, it follows that $M_{n,p} = \mathcal{O}(pb_n)$ which goes to $0$ as $n \rightarrow \infty$. The conditions for applying the quantitative central limit theorem are analogous to those in  Appendix~\ref{proof:first-div-dist} and can be verified in a straightforward way.

To deal with $R_{n,p}$, we rewrite $\bb{a}(t_i) = \bb{\eta}(t_i) + \bb{V}_1^{1/2}(t_i) \bb{\xi}(t_i)$ through application of Taylor's expansion of the normalized errors about the origin. Hence,
\begin{equation*}
    R_{n,p} = \dfrac{1}{n}\sum_{i=1}^n \bb{K}(t_i) \left[ \dfrac{\bb{I}_p}{\Vert \bb{\eta}(t_i)\Vert } - \dfrac{\bb{\eta}(t_i) \bb{\eta}(t_i)\tr}{\Vert \bb{\eta}(t_i)\Vert^3 } \right] \bb{K}(t_i) + \mathcal{O}(\sqrt{pb_n}),
\end{equation*}
where we use the fact that $\E\left[ \bb{\xi}(t_i) \mid \Gcal_{t_i-1} \right] = 0$ by independence of the errors and the filtration $\{ \Gcal_t : t\in \Tcal\}$ and mean-zero property of the errors. Note that, the matrix 
\begin{equation*}
    \Vert \bb{\eta}(t_i) \Vert^2 \bb{I}_p - \bb{\eta}(t_i)\bb{\eta}(t_i)\tr
\end{equation*}
is positive definite for all $n, p$ and $\bb{q} \notin \{\bb{Y}(t_i) : t_i \in \Tcal_n \}$. Therefore,
\begin{equation*}
    \left\Vert \nabla^2 \Mcal_{\bb{u},n}^{(p)}(\bb{q}) - \bb{\Psi}_t(\bb{q}_0) \right\Vert = \mathcal{O}\left(pb_n, \sqrt{pb_n} \right),
\end{equation*}
and the precise result in~\ref{F1-prop:second-div-mat} follows readily since $pb_n \rightarrow 0$, as $n\rightarrow \infty$.


\section{Simulation study}
\label{Sec:simulation}

In this section, we focus on the finite-sample performance of the proposed spatio-temporal quantile estimation methodology in two distinct simulation frameworks. The first takes place in a full-fledged spatio-temporal setting, where the locations are treated in terms of a multivariate time series framework, without explicitly incorporating information on the spatial coordinates. In contrast, the second case has been designed so as to draw on the inter-connectedness of spatial coordinates to emulate typical features in spatio-temporal data. Both setting were chosen upon careful consideration for illustrating the overall potential and efficacy of the proposed nonparametric quantile estimation as part of this paper. In this exercise, in order to evaluate efficiency, we use two popular metrics -- the mean absolute error (MAE) and the mean absolute percentage error (MAPE).

\subsection{Simulation setup 1}

As before, we denote by $p$ the number of spatial points, and let $s_1,\hdots,s_p$ denote the positions (or locations) of these points. In this setting, we assume that the exact coordinates of the locations are unknown. Therefore, to simplify the computations, we consider a regular set of the form $\{1,2,\hdots,p\}$ to denote the set of the spatial points. Throughout this experiment, data are generated for up to $T=300$ time points. The covariate series $\{\bb{X}(t)\}_{1\leqslant t\leqslant T}$ is assumed to be bivariate and we simulate it following a vector-valued autoregressive (VAR) series of order 1. In particular, $\bb{X}(t)$ is assumed to follows VAR(1) process with mean $\bb{0}$, satisfying the equation
\begin{equation*}
        \bb{X}(t) = \begin{bmatrix}
          0.2 & 0.1\\
          -0.3 & 0.4
        \end{bmatrix}
        \bb{X}(t-1) + \begin{bmatrix}
          w_{1t}\\
          w_{2t}
        \end{bmatrix},
\end{equation*}
where $w_{1t}, w_{2t}$ are uncorrelated white noise process. We shall use $\bb{X}_1(t)$ and $\bb{X}_2(t)$ to denote the two coordinates of the vector.
    
Next, the mean function associated with the response variable $\bb{Y}(t)$ (see \eqref{F1-eqn:setup} for reference) is assumed to take the form
\begin{equation*}
        \bb{\mu}(\bb{X}(t)) = \begin{bmatrix}
          \alpha_1 + \beta_{1,1} t + \beta_{2,1}\sin(2\pi \bb{X}_1(t)) + \beta_{3,1}\cos(2\pi \bb{X}_2(t)) \\
          \alpha_2 + \beta_{1,2} t + \beta_{2,2}\sin(2\pi \bb{X}_1(t)) + \beta_{3,2}\cos(2\pi \bb{X}_2(t)) \\
          \vdots \\
          \alpha_p + \beta_{1,p} t + \beta_{2,p}\sin(2\pi \bb{X}_1(t)) + \beta_{3,p}\cos(2\pi \bb{X}_2(t))
    \end{bmatrix}.
\end{equation*}

Note that the above is a nonlinear function in $\bb{X}(t)$, and this formulation is useful to understand if the specified nonparametric method is effective in a general scenario. To generate the intercept terms ($\alpha_i$) in the above expression, for $i = 1, 2, \dots, p$, we use the standard normal distribution. The $\beta_{1,i}$ coefficients indicate a linear trend pattern, and they are generated independently from uniform distribution between $-1$ and $1$. Finally, the parameters $\beta_{2,i}$ and $\beta_{3,i}$, for all $i$, are obtained independently from a uniform distribution in the interval $(-20, 20)$. Note that these choices ensure a greater degree of dependence on $\bb{X}(t)$, and are helpful to assess the accuracy of the proposed technique. 

For the covariance structure in the model \eqref{F1-eqn:setup}, we consider  
\begin{equation*}
    \bb{\Sigma}(\bb{X}(t)) = 0.1 \left\| \bb{\mu}(\bb{X}(t))\right\| \begin{bmatrix}
        \exp\{-0.1(s - s\dash)^2\}
    \end{bmatrix}_{s, s\dash = 1}^p,
\end{equation*}
the second term indicating a matrix with $(s,s')^{th}$ element given by the exponentially decaying covariance function. We point out that this specification maintains the signal to the noise ratio close to $0.1$ throughout the process.

Based on the above simulation setup, we generate $B = 1000$ datasets, each with $300$ time-points and $10$ spatial coordinates. Then, using Monte Carlo simulation, we can approximate the expectation in \eqref{F1-eqn:parameter} for any choice of $\bb{x}$ and $\bb{u}$, and, therefore, are able to optimize the objective function given in \eqref{F1-eqn:parameter} to obtain $Q(\bb{u} ,\bb{x})$. To assess the accuracy of the proposed technique, we consider this value as the true quantile which we wish to estimate. Now, we randomly choose $100$ datasets among those $1000$, and for each of them, we compute the proposed estimator $\bbhat{q}_n(\bb{u} ,\bb{x})$ as in \eqref{F1-eqn:estimator}. The errors are computed according to different sizes of the dataset, in terms of the number of time-points (denoted by $t$ in the next table). Corresponding MAE and MAPE values are summarized in \Cref{tab:sim-S1-MAE}. Here, we choose a Gaussian kernel over the temporal and spatial domain, and restrict our attention to the choice of $\bb{u}$ as $(2\tau - 1)\bb{1}_p/\sqrt{p}$, where $0 < \tau < 1$ is the quantile of interest.

\begin{table}[ht]
\centering
\caption{MAE (MAPE in parentheses) in estimating the $\bb{u}$-quantile at different time-points ($t$), where $\bb{u}=(2\tau - 1)\bb{1}_p/\sqrt{p}$, under simulation setup 1.}
\label{tab:sim-S1-MAE}
\begin{tabular}{cccccc}
  \toprule
 $\tau$ & $t=10$ & $t=50$ & $t=150$ & $t=250$ & $t=290$ \\ 
  \midrule
  0.50 & 3.93 (10.75) & 0.69 (3.72) & 1.23 (4.04)  & 2.46 (3.42) & 3.65 (1.24) \\ 
  0.75 & 5.16 (15.97) & 2.12 (4.29) & 1.39 (5.17) & 3.72 (2.71) & 4.21 (1.87) \\ 
  0.80 & 5.26 (20.44) & 2.24 (4.58) & 1.46 (5.20) & 4.07 (2.61) & 4.36 (2.07) \\ 
  0.85 & 5.50 (22.96) & 2.50 (5.12) & 1.63 (5.35) & 4.09 (2.52) & 5.20 (2.45) \\ 
  0.90 & 7.83 (25.44) & 3.07 (9.03) & 2.91 (6.89) & 4.78 (2.97) & 6.56 (3.37) \\ 
   0.95 & 14.22 (27.87) & 3.23 (12.89) & 4.32 (10.70) & 6.50 (4.09) & 9.91 (4.86) \\ 
   \bottomrule
\end{tabular}
\end{table}

From the above table, we can observe that the percentage errors reduce substantially for larger sizes of the dataset. In estimating the median ($\alpha=0.5$), our proposed method makes only about 1.2\% error if 290 observations per time series are allowed. The MAPE typically remains less than 3\% for moderate quantiles, and for more than 250 observations per spatial point. The errors are slightly higher for more extreme quantiles. For instance, at $\alpha=0.95$, our technique incurs around $10$ to $13$ percent error on average for moderately sized datasets, and that error reduces to only $4$ to $5$ percent on average for $t\geqslant 250$. 

In order to demonstrate the predictive performance of the method, we use the observations for $t\leqslant 290$ as the training set, and then forecast the quantiles for the next time-points. The MAE and MAPE thus computed, across 100 different datasets, are reported in \Cref{tab:sim-S1-pred-MAE}. We can see that the errors vary between 3\% to 9.2\%, and stays around 4 to 5 percent in most of the cases. The method is robust in predicting 1 to 5 steps ahead. For different quantiles though, we see slightly different behaviors. For central quantiles, typically the errors are marginally higher, whereas for the extreme quantiles, our approach registers good accuracy with the MAPE of around $3$ to $4.5$ percent.

\begin{table}[ht]
\centering
\caption{MAE (MAPE in parentheses) in predicting the $\bb{u}$-quantile, where $\bb{u}=(2\tau - 1)\bb{1}_p/\sqrt{p}$, for simulation setup 1.}
\label{tab:sim-S1-pred-MAE}
\begin{tabular}{cccccc}
  \toprule
   $\tau$ & $1$-step ahead & $2$-step ahead & $3$-step ahead & $4$-step ahead & $5$-step ahead \\ 
  \midrule
   0.50 & 11.653 (6.361) & 11.876 (6.496) & 9.305 (4.997) & 16.723 (9.18) & 9.57 (5.16) \\ 
   0.75 & 10.686 (5.776) & 10.496 (5.666) & 7.003 (3.693) & 14.093 (7.704) & 7.489 (3.999) \\ 
   0.80 & 10.129 (5.478) & 9.882 (5.318) & 6.473 (3.4) & 13.413 (7.322) & 7.012 (3.736) \\ 
   0.85 & 10.689 (5.839) & 9.342 (5.048) & 5.879 (3.076) & 12.588 (6.857) & 6.481 (3.449) \\ 
   0.90 & 9.847 (5.362) & 9.722 (5.287) & 6.009 (3.184) & 11.527 (6.261) & 5.835 (3.108) \\ 
   0.95 & 8.735 (4.666) & 8.166 (4.387) & 5.493 (2.876) & 11.629 (6.331) & 5.855 (3.095) \\ 
   \bottomrule
\end{tabular}
\end{table}

\subsection{Simulation setup 2}

In the second simulation scenario, unlike the previous multivariate setup where only partial spatial information is assumed to be known, we focus on a spatio-temporal framework to generate the data where the actual spatial information is used in an explicit manner. We follow the work of \cite{das2017analyzing}, where the authors considered the spatio-temporal estimation problem from a Bayesian paradigm.  

To develop this model, $p=15$ spatial locations $\bb{s}_1, \dots \bb{s}_{15}$ are first generated uniformly in the unit square $[0, 1]$. Thus, $\Scal_p = \{ \bb{s}_1, \dots \bb{s}_{15}\}$, and $\bb{s}_i = (s_{i1}, s_{i2})$ with $0 \leqslant s_{i1}, s_{i2} \leqslant 1$. Next, we take $200$ equidistant points in the interval $[0, 1]$ as the time-points. In other words, $\Tcal_n = \{ t_1, t_2, \dots t_n \}$ with $t_1 = 0, t_n = 1$ and $(t_i - t_{i-1}) = 1/(n-1)$ for all $i = 2, \dots n$.
    
We now assume that $\bb{X}(t) = t$ for all $t \in \Tcal = [0, 1]$. This can be construed as a linear trend function. The conditional quantile-curve of level $\tau \in [0, 1]$, conditioned on the temporal variable $t$ and spatial variable $\bb{s}$, takes the form
\begin{equation}\label{eqn:quantile-curve-setup2}
    Q(\tau \mid t, \bb{s}) = t \xi_1(\tau, \bb{s}) + (1-t) \xi_2(\tau, \bb{s}),
\end{equation}
so that
\begin{equation*}
    \xi_1(\tau, \bb{s}_i) = \left( 1 - \dfrac{s_{i1} + s_{i2}}{2} \right)\tau^2 + s_{i1} \dfrac{\log(1+\tau)}{2\log 2} + \dfrac{s_{i2}}{2}\tau^3
\end{equation*}
and
\begin{equation*}
    \xi_2(\tau, \bb{s}_i) = (1 - s_{i2}^2)\sin\left(\tau \pi /2 \right) + s_{i2} \dfrac{(e^{\tau} - 1)}{(e - 1)}.
\end{equation*}

Owing to the inverse relationship between the quantile and the distribution function, we can generate observations from the above model by simulating uniform $[0, 1]$ random variables and evaluating the quantile function at those generated values. More relevant details of this approach can be found in~\cite{das2017analyzing}.

Akin to the earlier scenario, we generate 100 datasets aimed at illustrating the finite sample behavior of the quantile function for the above case. Then, the accuracy of the model is computed by checking the errors of these estimates with respect to the true quantile given by~\eqref{eqn:quantile-curve-setup2}. In \Cref{tab:sim-S2-MAE}, we demonstrate the mean absolute errors (MAE) between the true quantile values and the fitted quantiles. The mean absolute percentage error (MAPE) is also shown in the brackets. Clearly, the maximum error is merely $16\%$ in case of extreme quantiles, for the quantiles closer to the median the error is less than $5\%$. Also, as expected, the errors tend to increase as we move towards the extreme quantile or the right hand of the temporal domain due to the boundary effects.

\begin{table}[ht]
\centering
\caption{MAE (MAPE in parentheses) in estimating the $\tau$-quantile at different time-points ($t$) under simulation setup 2.}
\label{tab:sim-S2-MAE}
\begin{tabular}{cccccc}
  \toprule
  $\tau$ & $t = 0$ & $t = 0.25$ & $t = 0.5$ & $t = 0.75$ & $t = 1$ \\ 
  \midrule
  0.50 &  0.129 (1.44) & 0.037 (0.47) & 0.019 (0.27) & 0.119 (2.07) & 0.388 (8.28) \\ 
  0.75 & 0.135 (1.51) & 0.033 (0.42) & 0.045 (0.66) & 0.211 (3.67) & 0.547 (11.66) \\ 
  0.8 & 0.155 (1.73) & 0.035 (0.44) & 0.056 (0.82) & 0.238 (4.14) & 0.587 (12.51) \\ 
  0.85 & 0.171 (1.92) & 0.037 (0.47) & 0.068 (1) & 0.263 (4.58) & 0.629 (13.4) \\ 
  0.9 & 0.185 (2.08) & 0.037 (0.48) & 0.08 (1.18) & 0.291 (5.06) & 0.67 (14.29) \\ 
  0.95 & 0.205 (2.29) & 0.042 (0.54) & 0.096 (1.42) & 0.323 (5.62) & 0.721 (15.37) \\ 
  0.99 & 0.314 (3.52) & 0.082 (1.05) & 0.111 (1.63) & 0.348 (6.06) & 0.758 (16.16) \\
   \bottomrule
\end{tabular}
\end{table}

We repeat the simulation exercise to measure the predictive performance of our proposed estimator as well. In this case, we use the first $90$ temporal points and $12$ locations as the training data, and use that to predict the behaviour of the quantiles for the rest of the $3$ locations for time-points $t = 0.91$ to $t = 1$. The summary measures MAE and MAPE are presented in \Cref{tab:sim-S2-MAE-pred}. It highlights that the error increases slowly for multiple step ahead predictions, with the percentage error in prediction being $17.5\%$ for extreme quantiles.

\begin{table}[ht]
\centering
\caption{MAE (MAPE given in parentheses) in predicting the $\tau$-quantile for different future time-points ($t$) under simulation setup 2 with $90$ time-points and $12$ locations in training set.}
\label{tab:sim-S2-MAE-pred}
\begin{tabular}{rlllll}
  \hline
 $\tau$ & $t = 0.91$ & $t = 0.92$ & $t = 0.93$ & $t = 0.94$ & $t = 0.95$ \\ 
  \hline
    0.5 & 0.343 (6.64) & 0.355 (6.93) & 0.367 (7.23) & 0.38 (7.54) & 0.393 (7.86)  \\ 
  0.75 & 0.158 (1.7) & 0.152 (1.64) & 0.146 (1.58) & 0.14 (1.52) & 0.134 (1.46)  \\ 
  0.8 & 0.352 (3.41) & 0.343 (3.34) & 0.335 (3.26) & 0.326 (3.19) & 0.318 (3.11)  \\ 
  0.85 & 0.695 (6.1) & 0.685 (6.03) & 0.675 (5.95) & 0.664 (5.87) & 0.654 (5.8) \\ 
  0.9 & 1.214 (9.69) & 1.204 (9.63) & 1.194 (9.56) & 1.184 (9.5) & 1.174 (9.43) \\ 
  0.95 & 1.922 (14.01) & 1.916 (13.97) & 1.909 (13.93) & 1.902 (13.89) & 1.896 (13.85) \\ 
  0.99 & 2.601 (17.65) & 2.599 (17.64) & 2.597 (17.63) & 2.596 (17.62) & 2.594 (17.61)  \\ 
  & & & & & \\
   \hline
   $\tau$ & $t = 0.96$ & $t = 0.97$ & $t = 0.98$ & $t = 0.99$ & $t = 1$ \\ 
  \hline
    0.5 & 0.406 (8.19) & 0.419 (8.53) & 0.433 (8.89) & 0.447 (9.25) & 0.461 (9.63) \\ 
  0.75 &  0.128 (1.4) & 0.123 (1.35) & 0.117 (1.29) & 0.112 (1.24) & 0.107 (1.19) \\ 
  0.8 &  0.309 (3.04) & 0.301 (2.97) & 0.293 (2.9) & 0.285 (2.83) & 0.277 (2.76) \\ 
  0.85 &  0.644 (5.72) & 0.634 (5.64) & 0.625 (5.57) & 0.615 (5.49) & 0.605 (5.42) \\ 
  0.9 &  1.164 (9.36) & 1.155 (9.3) & 1.145 (9.23) & 1.135 (9.17) & 1.126 (9.11) \\ 
  0.95 &  1.889 (13.81) & 1.882 (13.77) & 1.876 (13.73) & 1.869 (13.7) & 1.863 (13.66) \\ 
  0.99 &  2.593 (17.6) & 2.591 (17.59) & 2.589 (17.58) & 2.588 (17.58) & 2.586 (17.57) \\ 
  \hline
\end{tabular}
\end{table}

As a concluding remark to this study, it is imperative to highlight that the first simulation setup imitates the data generating mechanism presented in this paper. Thus, the results present in simulation setup 1 corroborate our theoretical findings. On the other hand, simulation setup 2 describes the data generation mechanism through the specification of the quantile function. Therefore, the generated data may not satisfy the model~\eqref{F1-eqn:setup}. Since this particular simulation illustrates actual spatio-temporal data, it is important to verify if the proposed estimator works equally well under this setup. The aforementioned simulation illustrates this point. Due to the potential violation of these model assumptions, the errors in simulation setup 2 are more than the errors for setup 1, though the percentage error remains at an acceptable level for non-extreme quantiles.



\subsection{Simulation for hypothesis testing}

In this section, we consider a short simulation study to assess the efficacy of the testing procedures discussed in \Cref{F1-sec:ht}. To that end, we consider the two following setups. 

\begin{enumerate}
    \item In the first setup, we choose the true quantile function as 
    \begin{equation*}
        Q(\tau, \bb{X}_t = x) = (-0.1 + x + 0.1 x^2 + \tau \abs{x}) \bb{1}_{10}; \ \tau \in [-1, 1]
    \end{equation*}
    \noindent where $\bb{1}_p$ is a $p$-variate vector with all entries equal to $1$. The values of the covariate $x$ were generated randomly from a normal distribution with mean $0$ and standard deviation $10$. To generate the observations $\bb{Y}_t$ at each timepoint $t = 1, 2, \dots n$ for $n = 100$, we generate $100$ standard uniform random variables and evaluate the quantile function at those points. Finally, random noise following a Gaussian distribution with mean $0$ and variance $\sigma^2$ are added to each coordinate of $\bb{Y}_t$ for each $t$ as random measurement errors. Clearly, in this case, the true quantile follows the null hypothesis of covariate homogeneity, if the error deviance $\sigma^2$ is low.
    
    \item In the second setup, we choose the true quantile function as     
    \begin{equation*}
        Q(\tau, \bb{X}_t = t) 
        = t \begin{bmatrix}
            \tau \\
            -\tau \\
            \tau^2 - \tau
        \end{bmatrix}.
    \end{equation*}
    \noindent Here also we apply a similar simulation procedure as in the previous step. In this case, the true quantile follows the null hypothesis of temporal homogeneity.
\end{enumerate}

\begin{figure}[!ht]
    \centering
    \includegraphics[scale=0.65]{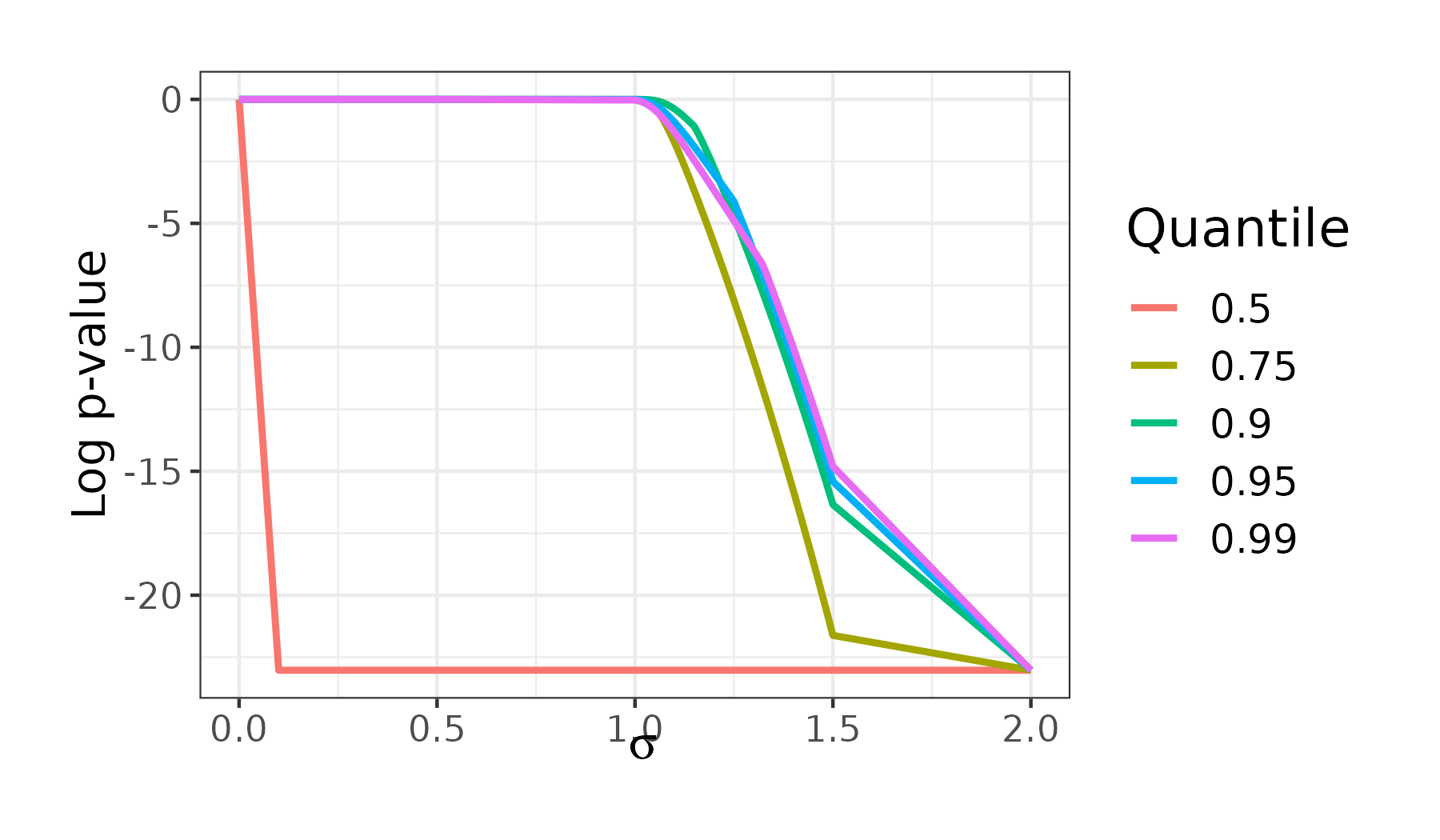}\\
    \includegraphics[scale=0.65]{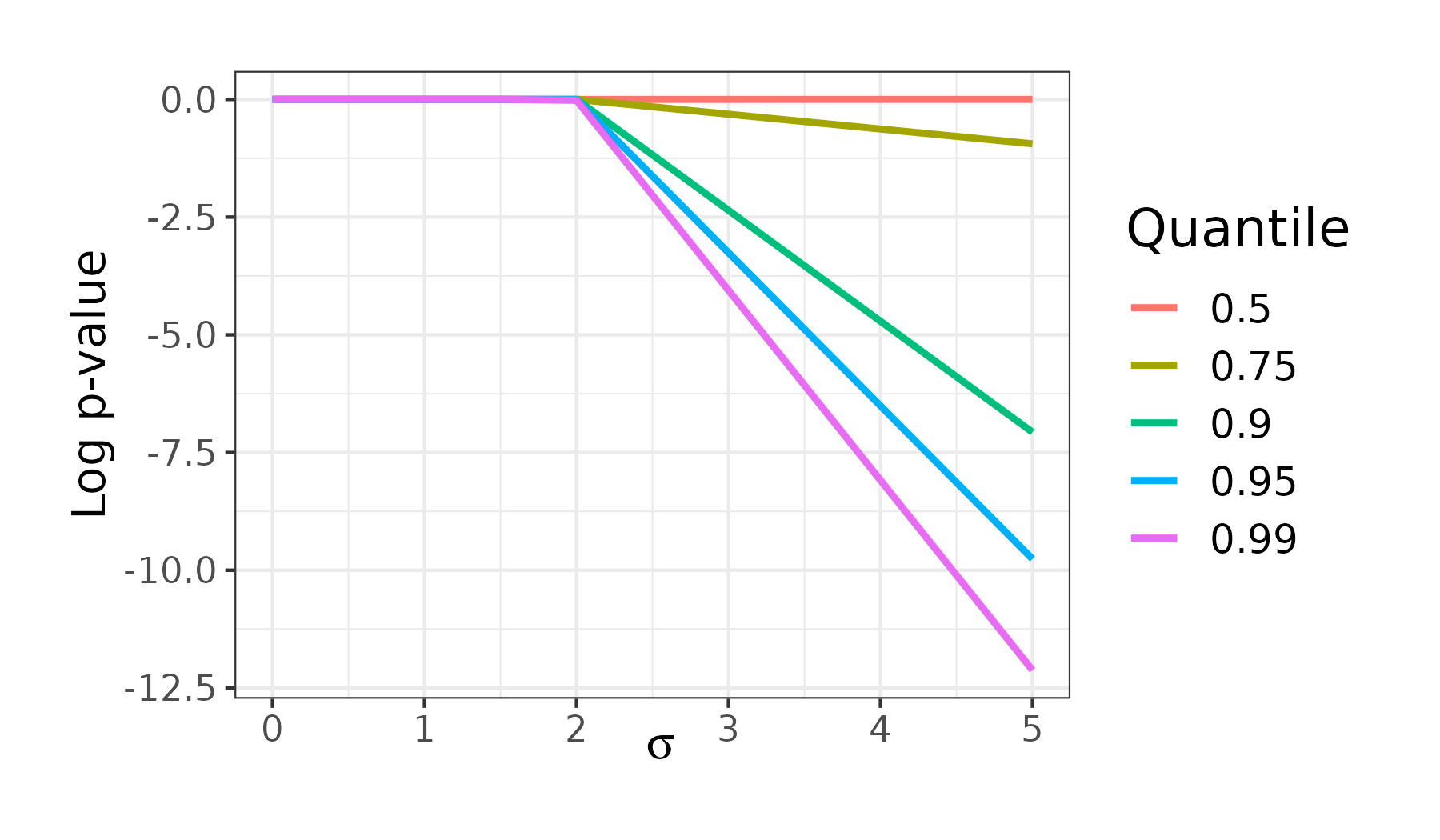}
    \caption{Evolution of $p$-values as a function of error variance from hypotheses testing on  spatial homogeneity \emph{(top)} and  temporal homogeneity \emph{(bottom)}.}
    \label{fig:ht-plot}
\end{figure}

For both of these setups, we change the error deviation by changing $\sigma$ and record the $p$-value as obtained by the hypothesis tests given in Section~\ref{F1-sec:ht}. The results are demonstrated in Figure~\ref{fig:ht-plot}. In both cases, for lower values of $\sigma$, the data is closely supported by the null hypothesis and hence we obtain a $p$-value of $1$. On the other hand, as $\sigma$ is increased, the data shows an indication of departure from homogeneity, as evident from the low $p$-values. For the test of spatial homogeneity, deviation from null is easily captured for the median, and it progressively becomes difficult for higher quantiles. On the other hand, the effect is reversed for the test of temporal homogeneity. This is due to the precise choice of the multivariate quantile $\bb{u} = (2\tau - 1)\bb{1}_p/\sqrt{p}$, as increasing variance affects the higher quantile curves more to deviate from a linear temporal trend across all coordinates.

\bibliographystyle{imsart-nameyear.bst} 
\bibliography{QRreferences.bib}